# Dynamic Task Fetching Over Time Varying Wireless Channels for Mobile Computing Applications

Aditya Dua, Dimitrios Tsamis, Nicholas Bambos, and Jatinder Pal Singh

**Abstract**

The processing, computation and memory requirements posed by emerging mobile broadband services require adaptive memory management and prefetching techniques at the mobile terminals for satisfactory application performance and sustained device battery lifetime. In this work we investigate a scenario where tasks with varied computational requirements are fetched by a mobile device from a central server over an error prone wireless link. We examine the buffer dynamics at the mobile terminal and the central server under varying wireless channel connectivity and device memory congestion states as variable sizes tasks are executed on the terminal. Our goal is to minimize the latency experienced by these tasks while judiciously utilizing the device buffering capability.We use a dynamic programming framework to model the optimal prefetching policy. We further propose a) a prefetching algorithm Fetch-or-Not (FON), which uses quasi-static assumption on system state to make prefetching decisions, and b) a prefetching policy RFON, which uses randomized approximation to the optimal solution thus obviating the need for dynamic online optimization and substantially reducing the computational complexity. Through performance evaluation under slow and fast fading scenarios we show that proposed algorithms come close to performance of the optimal scheme.

## I. Introduction

The advent of portable devices with wireless communication capability (e.g., PDAs, mobile phones) has provided great impetus to mobile computing applications. A broad spectrum of wireless broadband services are being offered to billions of users across the globe today. Some of these include location based services, streaming of compressed media (e.g. video) to mobile users, distributed execution of parallelizable computational tasks over

Aditya Dua is with Qualcomm Inc., 657 Campbell Technology Pkwy, Campbell CA 95008. Email: adua@qualcomm.com

Dimitrios Tsamis and Nicholas Bambos are with Department of Electrical Engineering, Stanford University, 350 Serra Mall, Stanford CA 94305. Email: {dtsamis,bambos}@stanford.edu

Jatinder Pal Singh is with Deutsche Telekom R&D Laboratories, 5050 El Camino Real, Los Altos, CA 94022, Email: jatinder.singh@telekom.com, and with Department of Electrical Engineering, Stanford University, 350 Serra Mall, Stanford CA 94305. Email: jatinder@stanford.edu

multiple cooperating mobile devices, etc. All these applications are executed on mobile terminals with limited processing power, battery life, and memory. Moreover, these terminals communicate with a central server/controller over error prone wireless links with fluctuating quality. Intelligent resource management and robust adaptation to variations in the wireless environment are therefore essential for optimizing the performance of mobile computing applications.

This paper focuses on dynamic task prefetching, adaptive processing, and memory management at mobile terminals (MT). Broadly speaking, our goal is to minimize the latency experienced by computational tasks, while judiciously utilizing the scarce memory resources available at the MT. More specifically, we are interested in a mobile computing scenario, where the MT sequentially fetches computational tasks from a central server (CS) over an error prone wireless link. It takes a random amount of time to transmit the task from the CS to the MT due to fluctuations in wireless channel quality. Further, it takes a random amount of time to complete the execution of each task at the MT due to resource contention with other competing tasks under execution at the MT.

The problem of joint buffer management and power control was addressed by Gitzenis and Bambos [2] in the context of client/server interaction for predictive caching. The focus of the work was to prefetch data over varying wireless channel using appropriate power levels and unlike the present work dynamic execution of application tasks and their server-end buffering was not addressed. The power aware prefetching problem was also studied by Cao in [3]. Dua and Bambos [4] examined buffer management for wireless media streaming, where the objective was to minimize buffer underflows to ensure smooth media playout, at the same time using the limited buffer at the mobile terminal in a careful manner. Buffer management for media streaming was also studied by Kalman et al. [5], Li et al. [6], etc. In other work on memory management in mobile computing scenarios, Ip et al. [7] proposed an adaptive buffer control scheme to prevent buffer overflows at MTs in a real-time object-oriented computing environment, and Yokoyama et al. [8] proposed a memory management architecture for implementing shared memory between the CS and the MT. To the best of our knowledge, *the latency vs. buffer tradeoff* in the mobile computing scenario delineated in this paper has not been addressed in the existing literature.

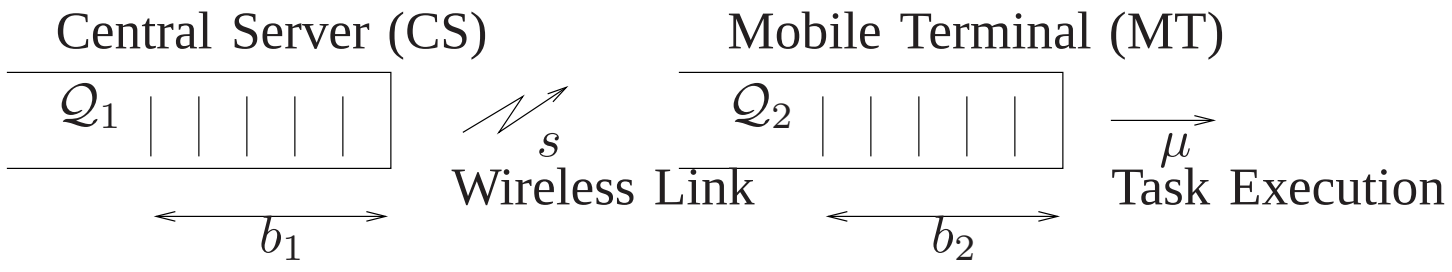

Fig. 1. System Model

For convenience, we will refer to a set of related tasks as an *application*. An application could be stand-alone, involving only the CS and the MT, or could also be part of a larger distributed computation involving multiple MTs. From the applications perspective, the best strategy is clearly to buffer all the tasks at the MT as quickly as possible. However, *memory is a limited and an expensive resource* at MTs and is shared by several applications (each one with its own set of tasks) which are executed concurrently. These applications could comprise of computational tasks fetched from a different CS or neighboring MT(s), or could also be locally generated system specific processes.

The MT needs to be "smart" in terms of the number of tasks it buffers locally for each application, because allocating a large chunk of memory to one application to improve its performance is likely to hurt the performance of other applications. From this perspective, the MT should request a new task from the CS as conservatively as possible. An exact analysis of the tradeoffs involved in this situation would involve modeling the dynamics of each application individually and considering the interactions induced between them by the shared memory resource. This holistic approach, however, is cumbersome (both analytically and computationally) and also not scalable. An alternative approach, which we adopt here, is to focus on the dynamics of one application (chosen arbitrarily) and model other applications as "background congestion" for this foreground application, and capture the coupling between them through a minimal set of parameters.

Since the foreground and background applications share a common limited resource — the memory — the background applications create *congestion* for the foreground application. This effect can be captured through a *congestion cost*. If there were no background applications, the congestion cost would be 0 and the entire memory could be dedicated to the foreground application. On the other hand, if there were a large number of background applications, the congestion cost would be quite large, and the foreground application would be allocated only a small chunk of the memory.

Based on the foregoing discussion, the dilemma faced by the MT in each decision epoch is the following: Fetch a task from the CS and possibly incur an additional congestion cost or not fetch a task and possibly increase the latency experienced by the application. To fetch or not to fetch?

We address the above trade-offs by modeling the problem within a dynamic programming framework [9]. We begin by the outlining the system model and the structural properties of the dynamic programming formulation in Section II. We discuss a special and useful instance of the optimization problem by employing a simple wireless channel and mobile terminal congestion models and establish the optimality of a switchover policy [10] in Section III. We then discuss in Sections IV and V, algorithms that use quasi-static and/or randomized approximations to the optimal solution. We evaluate the performance of the proposed prefetching algorithms in Section VI and conclude the work in Section VII.

## II. System Model

The mobile computing scenario we are interested in studying can be abstracted as a controlled two-queue tandem network, as depicted in Fig. 1. We assume that time is divided into identical time slots. In the figure, $\mathcal{Q}_1$ represents the queue at the central server (CS) and $\mathcal{Q}_2$ represents the queue at the mobile terminal (MT). The MT is interested in running an *application* assigned to it by the CS. An application comprises of a sequence of related *tasks* to be executed by the MT. A task includes a set of instructions to be processed by the MT, along with any auxiliary data that may be needed to process the instructions. For convenience, we assume that each task can be encoded in a single network packet to be transmitted over the air. We will therefore use the words task and packet interchangeably throughout the paper.

We note that the focus of this paper is on the problem formulation aspects and parsimonious mathematical modeling of a complex stochastic system. The two queue tandem formulation presented here is novel in the context of mobile computing scenarios. Also, the two queue tandem is a very hard control problem to analyze, as is evident from [11]–[13] and other similar literature. Our work is a first attempt to develop systematic approximations to the optimal control policies associated with this class of problems. We leverag these approximations and other provable properties of the model to develop practical algorithms.

### *A. Fluctuating wireless channel*

At most one packet can be transmitted over the fluctuating wireless communication link from the CS to the MT in a time slot. At the beginning of every time slot, the MT has the option of fetching a task from the CS over this error-prone wireless link. We model the variations in wireless channel reliability with a time homogeneous finite state Markov chain (FSMC). This model has been widely employed in the wireless communications and networking literature (see [14] and references therein). An FSMC channel model is characterized by a state-space $\mathcal{J} = \{1, \ldots, J\}$ and an associated $J \times J$ state-transition matrix $P$. Channel state transitions are assumed to occur at the end of each time slot. In particular, if the channel state at the beginning of the current time slot is $j$, it will transition to $j'$ at the end of the current time slot with probability (w.p.) $P(j, j')$. With the FSMC, we associate a mapping $s : \mathcal{J} \mapsto [0, 1]$, such that $s(j)$ denotes the probability of successful packet transmission if the MT chooses to fetch a packet from the CS when the channel is in state $j$.

### *B. Time varying congestion at the MT*

We also model the state of the processor at the MT as an FSMC. The FSMC is characterized by a state-space $\mathcal{M} = \{1, \ldots, M\}$ and a state-transition matrix $Q$. Associated with the Markov chain is a mapping $\mu : \mathcal{M} \mapsto (0, 1]$, such that the expected time to process a task when the processor is in state $m$ is $1/\mu(m)$ time slots. In other words, the task execution time in state $m$ is a geometrically distributed random variable (R.V.) with parameter $\mu(m)$. This model captures the randomness in task execution times at the MT, which can be attributed to two reasons:

1) Variable sized tasks*,
2) Contention for the shared processor at the MT (shared with tasks from other applications being executed at the MT).

*The "size" of a task refers to the computational resources it requires. Two tasks encoded in identical sized network packets can have very different computational requirements.

### C. System state and costs

Given the above mathematical model, the two queue tandem network can be completely described at any instant by the backlogs of the two queues, the channel state, and the processor state. More formally, let $\mathbf{x} = (b_1, b_2, j, m)$ denote the *state* of the system, where $b_1$ is the number of remaining tasks at the CS (in queue $\mathcal{Q}_1$) for the application of interest (foreground application), $b_2$ is the number of tasks waiting to be processed at the MT (in queue $\mathcal{Q}_2$), $j \in \mathcal{J}$ is the state of the wireless link from the CS to the MT (associated with probability of successful transmission $s(j)$), and $m \in \mathcal{M}$ is the state of the processor at the MT (associated with average task execution time $1/\mu(m)$). Also, for ease of notation, define $\mathbf{b} \triangleq (b_1, b_2)$, so that $\mathbf{x} = (\mathbf{b}, j, m)$.

A cost of 1 unit per task is incurred for every time slot that a task spends waiting in $\mathcal{Q}_1$, and a cost of $c \geq 1$ units per task is incurred for every time slot that a task spends in $\mathcal{Q}_2$.

*Remark 1 ($c$ captures the congestion vs. latency tradeoff):* The parameter $c$ represents the *congestion cost* as experienced by the foreground application at the MT. If $c$ is small, the MT is likely to fetch tasks from the CS as quickly as possible in order to reduce the overall latency experienced by the application. On the other hand, if $c$ is large, the MT is unlikely to fetch a task from the CS until its local buffer $\mathcal{Q}_2$ is empty, thereby resulting in higher latency for the application. The parameter $c$ therefore captures the tradeoff between the congestion cost and the latency cost. More importantly, $c$ captures the coupling between the background and foreground applications in a parsimonious fashion, without explicitly modeling the former. A well chosen value for $c$ ensures that the MT requests tasks judiciously from the the CS — infrequently enough to prevent buffer overflow at the MT (resulting in potential disruption of other applications competing for the shared MT resources) and frequently enough to prevent buffer underflows (resulting in potential processor under-utilization). Thus, $c$ is a critical design parameter and determines the operating point of the system. We will examine system performance as a function of $c$ via simulations in Section VI.

*Remark 2 (Buffering at the CS is not free):* Typically, availability of memory at the CS will not be a bottleneck in a real system. Then why should tasks queued at the CS incur a buffering cost of 1 unit per time slot in our model? This cost creates a *backlog pressure*, which drives down the overall latency experienced by the foreground application. To see this argument clearly, consider a scenario where tasks queued at the CS do not incur a backlog cost. Clearly then, in order to minimize overall buffering costs, the MT will request a new task only when it has finished processing the currently executing task. Since packet transmission times over the wireless link are random, the consequence would be potential under-utilization of the processor at the MT (especially in the absence of competing background applications), which is bad from an application latency perspective. A non-zero buffering cost at the CS prevents such a scenario from arising. If the MT requests tasks from the CS too infrequently in order to reduce buffering costs, the backlog pressure at the CS starts building up, which eventually forces the MT to request a task, and thereby drives down the application latency cost. Since any non-zero holding cost at the CS will suffice for generating the requisite backlog pressure, we set it to a normalized value of 1, without any loss of generality.

| New state | Transition probability |
|---|---|
| $(b_1, b_2-1, j', m')$ | $\mu(m)P(j,j')Q(m,m')$ |
| $(b_1, b_2, j', m')$ | $[1-\mu(m)]P(j,j')Q(m,m')$ |

TABLE I
POSSIBLE STATE TRANSITIONS AND ASSOCIATED TRANSITION PROBABILITIES IF A POLICY $\pi$ SELECTS ACTION $\overline{\text{FE}}$ IN SYSTEM STATE $(b_1, b_2, j, m)$, ASSUMING $b_2 > 0$.

| New state | Transition probability |
|---|---|
| $(b_1-1, b_2, j', m')$ | $s(j)\mu(m)P(j,j')Q(m,m')$ |
| $(b_1, b_2-1, j', m')$ | $s(j)\mu(m)P(j,j')Q(m,m')$ |
| $(b_1-1, b_2+1, j', m')$ | $s(j)[1-\mu(m)]P(j,j')Q(m,m')$ |
| $(b_1, b_2, j', m')$ | $[1-s(j)][1-\mu(m)]P(j,j')Q(m,m')$ |

TABLE II
POSSIBLE STATE TRANSITIONS AND ASSOCIATED TRANSITION PROBABILITIES IF A POLICY $\pi$ SELECTS ACTION FE IN SYSTEM STATE $(b_1, b_2, j, m)$, ASSUMING $b_1, b_2 > 0$.

### D. Actions and system dynamics

In our model, we assume that given the initial system state at time $t = 0$, no further tasks arrive to queue $\mathcal{Q}_1$ at the CS. Thus, the eventual state of the system when all tasks have been executed is $(0, 0, j, m)$, for some $j \in \mathcal{J}, m \in \mathcal{M}$, i.e., both queues will eventually be empty[†]. In other words, all states of type $(0, 0, j, m)$ are *terminal states* for the system. Given the system state $\mathbf{x}$ at the beginning of a time slot, the MT has to choose one of two actions:

1) FE: Fetch a task from the CS, or
2) $\overline{\text{FE}}$: do not fetch a task from the CS.

Our objective is to determine the *optimal policy*, or the optimal sequence of actions (FE or $\overline{\text{FE}}$) which drive the system from any given initial state to one of its terminal states, while incurring the lowest possible expected cost. We first formally define the notion of a policy.

*Definition 1:* A policy $\pi$ is a mapping $\pi : \mathbb{Z}_+ \times \mathbb{Z}_+ \times \mathcal{J} \times \mathcal{M} \mapsto \{\text{FE}, \overline{\text{FE}}\}$, which assigns one of the two possible actions (FE or $\overline{\text{FE}}$) to each system state $\mathbf{b}$.

The possible state transitions along with the associated state transition probabilities when policy $\pi$ chooses action $\overline{\text{FE}}$ in state $(b_1, b_2, j, m)$, assuming $b_2 > 0$, are tabulated in Table I. Similarly, the possible state transitions and associated probabilities when $\pi$ selects action FE in state $(b_1, b_2, j, m)$, assuming $b_1, b_2 > 0$, are listed in Table II. The state transitions are similarly described for the boundary cases $b_1 = 0,\ b_2 > 0$ and $b_1 > 0,\ b_2 = 0$.

[†]The case of stochastic task arrivals to the queue at the CS ($\mathcal{Q}_1$) can be easily studied in our framework, if the arrival process can be modeled as a discrete time Markov chain. For example, i.i.d. Bernoulli arrivals and correlated bursty arrivals fall into this category. Incorporating dynamic packet arrivals complicates the description and analysis of the system model, without providing significant additional insight into the system behavior. We therefore chose a "buffer draining model" for our exposition.

### E. Dynamic programming (DP) formulation

Given the system dynamics in Section II-D, we are interested in computing the optimal policy $\pi^\star$, which minimizes the total expected cost incurred in reaching a terminal state $(0, 0, \star, \star)$, starting in any state $\mathbf{x} = (\mathbf{b}, j, m)$. This is a *stochastic shortest path* (SSP) problem and is amenable to solution in a DP framework. Denoting by $V(\mathbf{b}, j, m)$ the *value function*, i.e., the expected cost incurred under the optimal policy $\pi^\star$ in reaching a terminal state, starting in state $(\mathbf{b}, j, m)$, we know from the theory of dynamic programming that $V(\cdot)$ satisfies the following *Bellman's equations* $\forall\ b_1, b_2 > 0$:

$$
\begin{aligned}
V(\mathbf{b}, j, m) = \min\{ & \sum_{j=1}^{J} \sum_{m=1}^{M} P(j, j') Q(m, m') [\mu(m) V(\mathbf{b} - \mathbf{e}_2, j', m') + \bar{\mu}(m) V(\mathbf{b}, j', m')], \\
& \sum_{j=1}^{J} \sum_{m=1}^{M} [s(j)\mu(m) V(\mathbf{b} - \mathbf{e}_1, j', m') + \bar{s}(j)\mu(m) V(\mathbf{b} - \mathbf{e}_2, j', m') + \\
& s(j)\bar{\mu}(m) V(\mathbf{b} - \mathbf{e}_1 + \mathbf{e}_2, j', m') + \bar{s}(j)\bar{\mu}(m) V(\mathbf{b}, j', m')]\} + \langle \mathbf{c}, \mathbf{b} \rangle,
\end{aligned}
\tag{1}
$$

where $\mathbf{b} = (b_1, b_2)$, $\mathbf{e}_1 = (1, 0)$, $\mathbf{e}_2 = (0, 1)$, $\mathbf{c} = (1, c)$, $\bar{s}(j) = 1 - s(j)$, $\bar{\mu}(m) = 1 - \mu(m)$, and $\langle \mathbf{c}, \mathbf{b} \rangle = b_1 + cb_2$. The first argument of $\min$ is the expected cost of choosing action $\overline{\text{FE}}$ in state $\mathbf{x}$, while the second argument is the expected cost of choosing action FE. Similar equations capture the boundary conditions $b_1 = 0$ or $b_2 = 0$. Finally, we have $V(0, 0, j, m) = 0\ \forall\ j, m$, i.e., all terminal costs are associated with zero cost.

## III. A Special and Insightful Case

The DP equations in (1) can be solved numerically to compute the optimal policy $\pi^\star$. Well developed numerical techniques like value iteration and policy iteration are available to solve the DP equations in an efficient manner [9]. In principle, the equations can be solved *offline*, and the optimal action for every possible system state can be stored in a lookup table (LuT) at the MT. In an *online* implementation, the MT simply looks at the current system state (backlogs of $\mathcal{Q}_1$ and $\mathcal{Q}_2$, channel state, and processor state) and extracts the optimal action (FE or $\overline{\text{FE}}$) from the LuT. Such an implementation is, however, fraught with the following difficulties:

1) The DP formulation presented above assumes a *fixed* value of $c$. In practice, the MT may wish to update the value of $c$, based on its observation of the processor utilization, channel conditions, etc., to drive the system to a desired operating point. Recall that $c$ captures the tradeoff between congestion and latency and hence determines the operating point of the two queue tandem. Thus, every time $c$ changes, the DP equations in (1) need to solved again, which is a computationally cumbersome task.

2) Even if computational complexity is not an issue, obtaining/estimating all the system parameters in real time is a non-trivial task. In particular, the MT may need an unacceptably long time to empirically estimate with sufficient accuracy the state transition matrices $P$ and $Q$, associated with the wireless channel and processor utilization, respectively. In a realistic setting, the MT can at best hope to estimate the instantaneous probability of successful transmission and instantaneous task execution speed (or the corresponding values averaged over a moving window).

Keeping the above implementation issues in mind, it is critical to devise a fetching algorithm which does not require frequent recomputation of the optimal policy from the DP equations, and also does not have unrealistic requirements in terms of the system parameters it needs to be cognizant of. To this end, we now turn our attention to a simple and insightful instance of the general system model described in Section II.

In particular, we introduce the following two *modeling reductions*:

1) *Wireless Channel*: The multi-state FSMC wireless channel is replaced by a two-state ON/OFF Bernoulli channel, which can be in ON state w.p. $s$ and in OFF state w.p. $1-s$ in very time slot, independent of its state in past and future time slots. When the channel is in ON state, a packet transmission over the channel is successful w.p. 1, and when the channel is in OFF state, a packet transmission over the channel fails w.p. 1. Using notation introduced in Section II-A, this i.i.d. channel model is characterized by state space $\mathcal{J} = \{1\}$ and state transition matrix $P = [1]$. Since $|\mathcal{J}| = 1$, the mapping $s : \mathcal{J} \mapsto [0, 1]$ reduces to a scalar $s$. Note that under the i.i.d. channel model, packet transmission times are geometrically distributed with mean transmission time $1/s$ slots.
2) *Congestion at the MT*: We assume that the expected time to process each task at the MT follows an i.i.d. geometric distribution with parameter $\mu$. Again, this is a special case of the multi-state FSMC model used in the previous section to model time varying congestion at the MT. In terms of the notation used in Section II-B, the simplified model is associated with the state-space $\mathcal{M} = \{1\}$ and state transition matrix $Q = [1]$. Since task processing times are now characterized by a single geometric distribution, the mapping $\mu : \mathcal{M} \mapsto [0, 1]$ reduces to the scalar $\mu$.

To summarize, the reduced system model is characterized by two key parameters: the probability of successful packet transmission over the wireless link from the CS to the MT (denoted $s$), and the average time needed to process each task at the MT (denoted $1/\mu$). Since these parameters are fixed for each instantiation of the reduced model (in contrast to the general model of Section II, where they were modeled as FSMCs with states denoted by $j$ and $m$, respectively), they need not be included in the description of the system state. Thus, the state associated with the reduced system model is two dimensional (in contrast to the four dimensional state of the general model), and is denoted by $\mathbf{b} = (b_1, b_2)$.

The action space associated with the reduced model remains unchanged, i.e., each state is associated with one of two possible actions, FE or $\overline{\text{FE}}$. The system dynamics can be described as in Section II-D. The value function now obeys the following Bellman's equations (obtained as a special case of (1), with $J = 1$ and $M = 1$):

$$V(\mathbf{b}) = \min\{\underbrace{\mu V(\mathbf{b}-\mathbf{e}_2) + \bar{\mu} V(\mathbf{b})}_{\text{Action } \overline{\text{FE}}}, \underbrace{s\mu V(\mathbf{b}-\mathbf{e}_1) + \bar{s}\mu V(\mathbf{b}-\mathbf{e}_2) + s\bar{\mu} V(\mathbf{b}-\mathbf{e}_1+\mathbf{e}_2) + \bar{s}\bar{\mu} V(\mathbf{b})}_{\text{Action FE}}\} + \langle \mathbf{c}, \mathbf{b} \rangle, \tag{2}$$

where $\bar{s} = 1 - s$, $\bar{\mu} = 1 - \mu$, and the rest of the notation is as defined in Section II-E. Finally, the terminal state associated with the above DP problem is $(0, 0)$, with $V(0, 0) = 0$.

*Remark 3 (Advantages of model reduction):* The modeling reductions introduced above help reduce the dimensionality of the problem from 4 to 2, which greatly facilitates analysis and algorithm design. Further, fetching algorithms designed on the basis of the reduced model do not suffer from the implementation hurdles discussed at the beginning of this section. We will revisit this claim in greater detail when we discuss algorithm design in the next section. Further, we will demonstrate via simulations that fetching algorithms based on the reduced model can closely match the performance of those based on the full fledged model of Section II.

Even though the DP equations associated with the reduced model cannot be solved in closed form, numerous interesting structural properties of the optimal solution can be established, which provide useful insights and intuition about the decision tradeoffs inherent in the problem, in addition to aiding low complexity algorithm design. We prove one such structural property below and illustrate a few others via numerical examples.

### A. Switchover property

We begin by providing the formal definition of a *switchover* type policy.

*Definition 2:* A policy $\pi$ is of *switchover* type if there exists a non-decreasing switchover curve $\psi : \mathbb{Z}_+ \mapsto \mathbb{Z}_+$ such that $\pi$ chooses action $\overline{\text{FE}}$ in state $(b_1, b_2)$ if $b_2 > \psi(b_1)$, and chooses action FE otherwise.

A switchover policy splits the two dimensional state-space into two distinct decision regions, one corresponding to each of the actions FE and $\overline{\text{FE}}$. The optimal policy of interest here, viz. $\pi^\star$, is of switchover type, in the sense of the above definition.

*Theorem 1:* The optimal policy $\pi^\star$ is of switchover type.

*Proof:* See Appendix VIII-A. ■

*Remark 4 (A similar result):* Theorem 1 is similar to a continuous time result proved by Rosberg et al. [11], where packet transmission times and task execution times were assumed to be exponentially distributed. However, the authors in [11] allowed for new task arrivals to $\mathcal{Q}_1$, and their objective was to compute the optimal policy which minimizes the infinite horizon discounted expected cost.

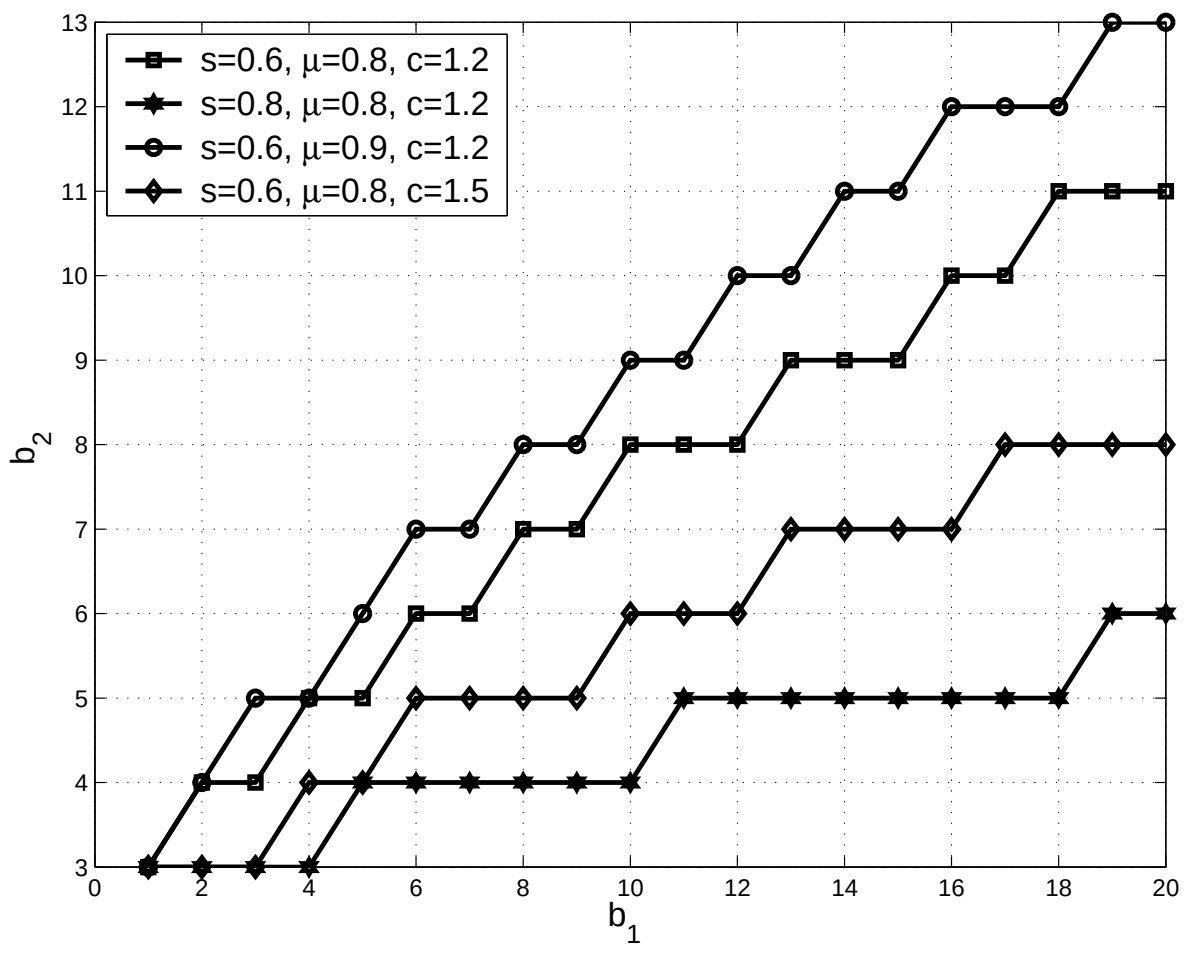

Fig. 2. Numerical Example 1

*Remark 5 (Load balancing effect):* The switchover property of $\pi^\star$ is intuitively quite appealing. When $b_1 \gg b_2$, i.e., the queue at the CS is much more loaded than the queue at the MT, the optimal action for the MT is FE. As the MT fetches more and more packets, $b_1$ starts decreasing and $b_2$ starts increasing, and the optimal decision eventually *switches over* from FE to $\overline{\text{FE}}$. Similarly, when $b_2 \gg b_1$, the optimal action is $\overline{\text{FE}}$. Thus, the MT stops fetching tasks for the CS and the size of $\mathcal{Q}_2$ relative to $\mathcal{Q}_1$ starts diminishing, until eventually the optimal action *switches over* from $\overline{\text{FE}}$ to FE. Thus, the switchover nature of the optimal policy induces a *load balancing* effect between the two queues $\mathcal{Q}_1$ and $\mathcal{Q}_2$.

We conclude this section by demonstrating some more structural properties of the optimal policy $\pi^\star$ via numerical examples.

*Numerical Example 1:* This example illustrates the behavior of the optimal policy $\pi^\star$ for different model parameters. Fig. 2 depicts the optimal switchover curve (computed from (2)) for different combinations of $s$, $\mu$, and $c$. In particular, we consider the following combinations: (0.6, 0.8, 1.2), (0.8, 0.8, 1.2), (0.6, 0.9, 1.2), and (0.6, 0.8, 1.5).

We make the following observations from the numerical examples:

1) The decision region for action FE gets smaller as $s$ increases, for fixed $\mu, c$. As the wireless channel becomes more reliable, the MT can afford to fetch tasks from the CS less aggressively, since fewer attempts are needed on an average to fetch a packet successfully.
2) The decision region for action FE grows bigger as $\mu$ increases, for fixed $s, c$. Since increasing $\mu$ decreases average task execution times, the MT has to fetch tasks more frequently from the CS in order to drive down application latency.
3) The decision region for action FE gets smaller as $c$ increases, for fixed $s, \mu$. A bigger $c$ indicates a higher

congestion cost for the foreground application, forcing the MT to be conservative in fetching tasks from the CS.

## IV. Dynamic Task Fetching Algorithms

In Section II, we developed a mathematical framework based on dynamic programming for studying dynamic task fetching algorithms for mobile computing scenarios. In Section III we briefly discussed an LuT based implementation of the optimal policy obtained by solving the DP equations. To surmount the practical difficulties associated with implementing the LuT based approach, we then turned our attention to a reduced but insightful version of the general model of Section II. The probability of successful packet transmission over the wireless link and the mean task execution time at the MT are fixed under the reduced model — assumptions which will clearly be violated in real life scenarios. Therefore, in this section we focus on designing task fetching algorithms for *dynamic* scenarios (i.e., time varying successful transmission probability and mean task execution times), while leveraging the simplicity of the *static* solution obtained from the analysis of the reduced model. We propose two algorithms:

- **Fetch-Or-Not (FON)**: This algorithm is based on the solution to the DP equations in (2).
- **Randomized Fetch-Or-Not (RFON)**: This algorithm is based on a randomized approximation (denoted by **RAND**) of the solution to the DP equations in (2). Thus, RFON does not actually need to solve any DP equations in real time.

Both FON and RFON are based on the notion of *quasi-static* decision making. Broadly speaking, a quasi-static decision algorithm bases its decision at every decision epoch on the solution to a static instance/snapshot of a dynamic system, where the static instance is constructed based on the *instantaneous operating point* of the system. As the instantaneous operating point of the system evolves from one epoch to another (governed by the inherent system dynamics), the parameters of the static snapshots used by the quasi-static algorithm change accordingly. This approach was used to great effect for developing downlink wireless packet scheduling algorithms by Dua et al. [15]. In the context of this paper, the instantaneous operating point refers to the two tuple $(s, \mu)$, i.e., the probability of successful transmission and the task execution rate at the MT.

### A. FON and RFON

The FON algorithm works as follows: Given the instantaneous estimates of $s$ and $\mu$ in time slot $t$, denoted by $\hat{s}(t)$ and $\hat{\mu}(t)$, the DP equations associated with the reduced model, viz. (2), are solved with $s = \hat{s}(t), \mu = \hat{\mu}(t)$, to select either decision FE or $\overline{\text{FE}}$ for the current backlog vector $\mathbf{b}(t) = (b_1(t), b_2(t))$. Based on the outcome of the decision (FE or $\overline{\text{FE}}$), the backlog vector changes to $\mathbf{b}(t+1) = (b_1(t+1), b_2(t+1))$. Estimates of the success probability and mean task execution time are also updated to $\hat{s}(t+1)$ and $\hat{\mu}(t+1)$ respectively, based on measurements made by the MT[‡].

The RFON algorithm is fairly similar to the FON algorithm, except that the decision in each time slot is based on a randomized approximation (RAND) to $\pi^\star$, instead of $\pi^\star$. Thus, RFON has lower computational complexity than FON because, unlike FON, it does not need to solve the DP equations in (2) in every time slot. We will devote the next section to studying the randomized approximation RAND.

Both FON and RFON algorithms are summarized in Table III.

| FON and RFON |
|---|
| In time slot $t$, |
| **Given** |
| estimate of probability of successful transmission $\hat{s}(t)$, |
| estimate of task execution rate $\hat{\mu}(t)$ |
| choice of congestion cost rate $\hat{c}(t)$, |
| **Compute** |
| $\pi^\star$ from (2) (for FON) or RAND (for RFON) with $s = \hat{s}(t)$, $\mu = \hat{\mu}(t)$, and $c = \hat{c}(t)$ |
| **Select** action FE/$\overline{\text{FE}}$ based on outcome of $\pi^\star$ (for FON) or RAND (for RFON) |
| **Update** $\hat{s}(t) \to \hat{s}(t+1)$, $\hat{\mu}(t) \to \hat{\mu}(t+1)$, and $\hat{c}(t) \to \hat{c}(t+1)$ |

TABLE III
DYNAMIC FETCHING ALGORITHMS: FON AND RFON

*Remark 6 ($c$ can be adapted to achieve a desired tradeoff):* In addition to dynamically estimating $s$ and $\mu$, the MT can vary/adapt the tradeoff parameter $c$ to achieve a desired tradeoff between application latency and local buffer congestion. Recall that under the system model of Section II (and its reduced version in Section III), $c$ was assumed to be fixed. We will not consider algorithms for dynamically adapting $c$ in this paper. Instead, we will simulate the performance of the algorithms FON and RFON with different (but fixed) values of $c$ to "sweep" tradeoff curves for the system.

## V. RAND: A RANDOMIZED APPROXIMATION TO $\pi^\star$

In this section, we develop a randomized approximation RAND to $\pi^\star$, based on the method of policy iteration [9]. Recall from Section IV that RAND is at the core of the dynamic fetching algorithm RFON.

Our first step is to analyze two "extreme" policies (under the assumptions of the reduced model of Section III):

1) $\pi_N$: Never fetches a task from the CS, until the buffer at the MT is empty, and
2) $\pi_A$: Continues to fetch tasks from the CS, until the buffer at the CS has been drained.

### A. The "never fetch" policy: $\pi_N$

Consider policy $\pi_N$ which *never* chooses to fetch a task from the CS, except when the buffer at the MT (queue $\mathcal{Q}_2$) is empty, i.e., $\pi_N$ selects action $\overline{\text{FE}}$ in all states $(b_1, b_2)$ with $b_2 > 0$. We are interested in computing the

[‡]A discussion of the estimation algorithms used by the MT is beyond the scope of this paper. For ease of presentation, we will assume that the MT can estimate both parameters accurately. However, it is important to note that the MT can only estimate instantaneous values of these parameters and not the underlying stochastic processes which drive the system dynamics.

expected cost incurred under $\pi_N$ in reaching terminal state $(0,0)$, as a function of the initial state $(b_1, b_2)$. Denoting this cost by $C_N(\mathbf{b})$, we have:

*Lemma 1:*

$$C_N(\mathbf{b}) = \left(\frac{\bar{\mu}}{\mu} + \frac{1}{s}\right)\frac{b_1^2}{2} + \frac{cb_2^2}{2\mu} + \frac{b_1 b_2}{\mu} + \left[\frac{1}{2}\left(\frac{\bar{\mu}}{\mu} + \frac{1}{s}\right) + \frac{(c-1)\bar{\mu}}{\mu}\right] b_1 + \frac{cb_2}{2\mu}.$$

*Proof:* See Appendix VIII-B. ■

### B. The "always fetch" policy: $\pi_A$

Now, in contrast to $\pi_N$, consider a policy $\pi_A$ which *always* chooses to fetch a task from the CS if available, i.e., it chooses the action FE in all states $\mathbf{b}$ with $b_1 > 0$. Again, we are interested in computing the expected cost incurred under $\pi_A$ in reaching terminal state $(0,0)$, as a function of the initial state $(b_1, b_2)$. We will denote the cost by $C_A(\mathbf{b})$. It is not possible to compute $C_A(\mathbf{b})$ in closed form; therefore, we approximately compute $C_A(\mathbf{b})$ from a *fluid caricature model*. We denote the corresponding cost in the fluid model by $C_A^f(\mathbf{b})$. The attributes of the fluid model are described in Appendix VIII-C. We have the following result for the fluid caricature model:

*Lemma 2:*

$$C_A^f(\mathbf{b}) = \begin{cases} \left(\frac{c}{\mu} - \frac{c-1}{s}\right)\frac{b_1^2}{2} + \frac{cb_2^2}{2\mu} + \frac{cb_1 b_2}{\mu} & ; \quad s \geq \mu \text{ or } s < \mu \text{ and } T_1 < T_0 \\ \frac{b_1^2}{2s} + \frac{cb_2^2}{2(\mu - s)} & ; \quad s < \mu \text{ and } T_1 \geq T_0, \end{cases}$$

where $T_0 \triangleq \frac{b_1}{s}$ and $T_1 \triangleq \frac{b_2}{\mu - s}$.

*Proof:* See Appendix VIII-C. ■

We are going to use $C_A^f(\mathbf{b})$ computed in Lemma 2 as an approximation to $C_A(\mathbf{b})$. We explore the efficacy of the approximation via a numerical example in Appendix VIII-C.

### C. Policy Iteration

*Policy iteration* is a well known numerical technique for solving Bellman's equations [9]. Given a feasible policy $\pi$, each iteration in policy iteration comprises of two steps:

1) *Policy evaluation*: In this step, the expected cost incurred under policy $\pi$, denoted $V_\pi(\mathbf{b})$, is evaluated $\forall\ \mathbf{b}$.

2) *Policy improvement*: In this step, the policy $\pi$ is "improved" to obtain a new policy $\pi'$. The improved policy $\pi'$ is computed as follows:

$$\pi'(\mathbf{b}) = \arg\min\{\underbrace{\mu V_\pi(\mathbf{b}-\mathbf{e}_2)+\bar{\mu}V_\pi(\mathbf{b})}_{\text{Action }\overline{\text{FE}}}, \underbrace{s\mu V_\pi(\mathbf{b}-\mathbf{e}_1)+\bar{s}\mu V_\pi(\mathbf{b}-\mathbf{e}_2)+s\bar{\mu}V_\pi(\mathbf{b}-\mathbf{e}_1+\mathbf{e}_2)+\bar{s}\bar{\mu}V_\pi(\mathbf{b})}_{\text{Action FE}}\}, \quad (3)$$

where $\pi'(\mathbf{b})$ denotes the action chosen by policy $\pi'$ (either $\overline{\text{FE}}$ or FE) in state $\mathbf{b}$.

The policy $\pi'$ is called a *one step improvement* of $\pi$. We are interested in computing one step improvements of the policies $\pi_N$ and $\pi_A$, defined in Sections V-A and V-B, respectively. We have already performed the policy evaluation step for both policies. In particular, we have $V_{\pi_N}(\mathbf{b}) = C_N(\mathbf{b})$ and $V_{\pi_A}(\mathbf{b}) \approx C_A^f(\mathbf{b})$.

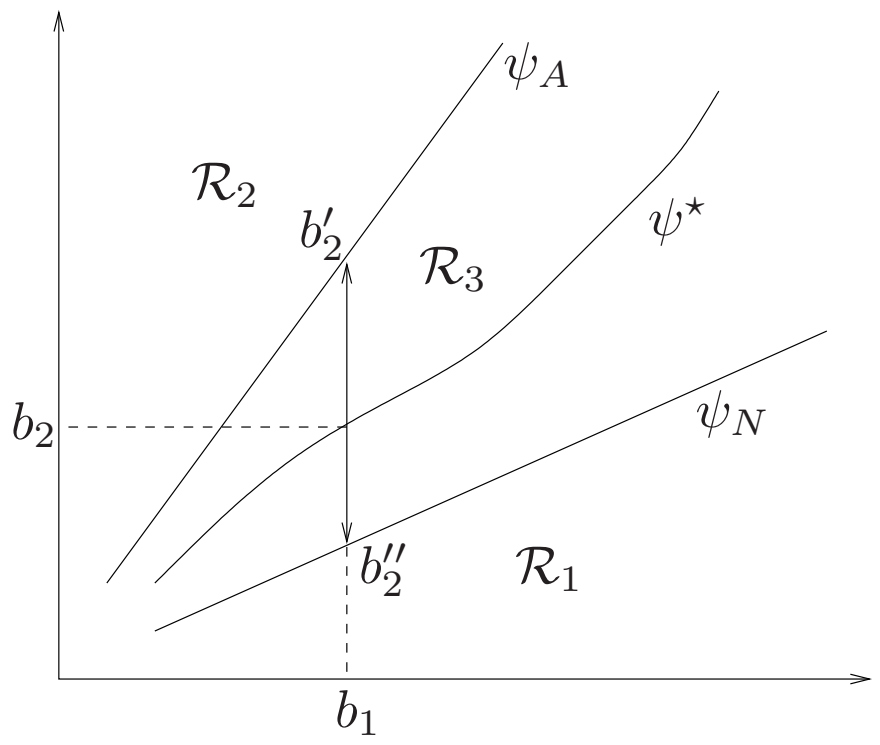

Fig. 3. Bounding the optimal switchover curve

Recall that $C_N$ and $C_A^f$ are quadratic functions of $b_1$ and $b_2$. We now therefore compute the one step improvement of an arbitrary policy $\pi$ with a quadratic cost of the form:

$$V_\pi(b) = \alpha_1 b_1^2 + \alpha_2 b_2^2 + \gamma b_1 b_2 + \beta_1 b_1 + \beta_2 b_2. \quad (4)$$

For convenience, define:

$$\begin{aligned} V_\pi^1(\mathbf{b}) &\triangleq V_\pi(\mathbf{b}+\mathbf{e}_1) - V_\pi(\mathbf{b}) \\ V_\pi^2(\mathbf{b}) &\triangleq V_\pi(\mathbf{b}+\mathbf{e}_2) - V_\pi(\mathbf{b}). \end{aligned}$$

The policy improvement equation (3) can be rewritten as

$$\pi'(\mathbf{b}) = \arg\min\{0, s\mu[V_\pi^2(\mathbf{b}-\mathbf{e}_2) - V_\pi^2(\mathbf{b}-\mathbf{e}_1)] + s[V_\pi^2(\mathbf{b}-\mathbf{e}_1) - V_\pi^1(\mathbf{b}-\mathbf{e}_1)]\} \quad (5)$$

It easily follows from (4) that

$$\begin{aligned} V_\pi^1(\mathbf{b}) &= 2\alpha_1 b_1 + \gamma b_2 + \alpha_1 + \beta_1 \\ V_\pi^2(\mathbf{b}) &= \gamma b_1 + 2\alpha_2 b_2 + \alpha_2 + \beta_2. \end{aligned} \tag{6}$$

Substituting (6) in (5) gives

$$\pi'(\mathbf{b}) = \arg\min\{0, \ell(\mathbf{b})\}, \tag{7}$$

where $\ell(\mathbf{b})$ is a linear function of $b_1$ and $b_2$. Note that the decision of $\pi'$ in state $\mathbf{b}$ is completely determined by the sign of $\ell(\mathbf{b})$. Since $\ell(\mathbf{b})$ is linear (i.e., of the form $a_1 b_1 + a_2 b_2 + a_3$ for some $a_1, a_2, a_3 \in \mathbb{R}$), the two dimensional state-space $(b_1, b_2)$ gets split into two distinct decision regions by a straight line, corresponding to the two decisions FE and $\overline{\text{FE}}$. In other words, policy $\pi'$ is of switchover type, in the sense of Definition 2. A little bit of algebra shows that $a_1 = \gamma - 2\alpha_1$, $a_2 = 2\alpha_2 - \gamma$, and $a_3 = \alpha_1 + (1 - 2\mu)\alpha_2 - \beta_1 + \beta_2 - (1 - \mu)\gamma$.

Based on the above analysis and the expressions for cost functions derived in Lemma 1 and Lemma 2, we can compute the switchover curves for $\pi'_A$ and $\pi'_N$, which are one step improvements of $\pi_A$ and $\pi_N$, respectively. We will refer to the switchover curves associated with $\pi'_A$ and $\pi'_N$ as $\psi_A$ and $\psi_N$, respectively.

## *D. The RAND policy*

It can be argued by contradiction that $\psi_A$ bounds $\psi^\star$ (the optimal switchover curve associated with $\pi^\star$) from above and $\psi_N$ bounds $\psi^\star$ from below, as depicted in Fig. 3. Thus, roughly speaking, we have bounded the optimal switchover curve in a "conical" region defined by $\psi_A$ and $\psi_N$. The cone splits the state-space into three distinct regions — $\mathcal{R}_1$, $\mathcal{R}_2$, and $\mathcal{R}_3$, as shown in Fig. 3. In region $\mathcal{R}_2$, which lies above $\psi_A$, the optimal action is $\overline{\text{FE}}$. In region $\mathcal{R}_1$, which lies below $\psi_N$, the optimal action is FE. Region $\mathcal{R}_3$, which is the interior of the cone, is a region of uncertainty. Our analysis tells us that $\psi^\star$ lies somewhere in region $\mathcal{R}_3$, but not exactly where.

We overcome the uncertainty in region $\mathcal{R}_3$ by employing a randomized decision policy. In particular, consider a state $\mathbf{b} = (b_1, b_2) \in \mathcal{R}_2$. We know that $\exists\ b'_2 \geq b_2$ such that the state $(b_1, b'_2)$ lies on the surface of the cone (on the switchover curve $\psi_A$) and the optimal decision is $\overline{\text{FE}}$ in all states $(b_1, y)$ with $y > b'_2$. Similarly, we know that $\exists\ b''_2 \leq b_2$ such that the state $(b_1, b''_2)$ lies on the surface of the cone (on switchover curve $\psi_N$) and the optimal decision is FE in all states $(b_1, y)$ with $y < b''_2$. If $(b_1, b_2)$ is closer to $\psi_A$ than $\psi_N$, then the optimal decision is more likely to be $\overline{\text{FE}}$, and if $(b_1, b_2)$ is closer to $\psi_N$ than $\psi_A$, then the optimal decision is more likely to be FE. In particular, for any state $(b_1, b_2) \in \mathcal{R}_3$, we will make a randomized decision based on the policy **RAND**, as described in Table IV.

*Numerical Example 2:* This example illustrates the bounding of the optimal switchover curve $\psi^\star$ in a "conical" region generated by the switchover curves $\psi_A$ and $\psi_N$, for two different sets of parameter values $s$, $\mu$, and $c$. The

| RAND |
|---|
| In state $(b_1, b_2)$, select action $\overline{\text{FE}}$ w.p. $\dfrac{b_2' - b_2}{b_2' - b_2''}$ |
| In state $(b_1, b_2)$, select action FE w.p. $\dfrac{b_2 - b_2''}{b_2' - b_2''}$ |
| $b_2' = \underset{(b_1,y)\in\psi_A}{\arg\min}\ \lvert y - b_2\rvert$ and $b_2'' = \underset{(b_1,y)\in\psi_N}{\arg\min}\ \lvert b_2 - y\rvert$ |

TABLE IV

RAND: A RANDOMIZED APPROXIMATION TO THE OPTIMAL POLICY $\pi^{\star}$.

results are depicted in Fig. 4. Observe that in both cases the conical region bounds $\psi^{\star}$ reasonably tightly.

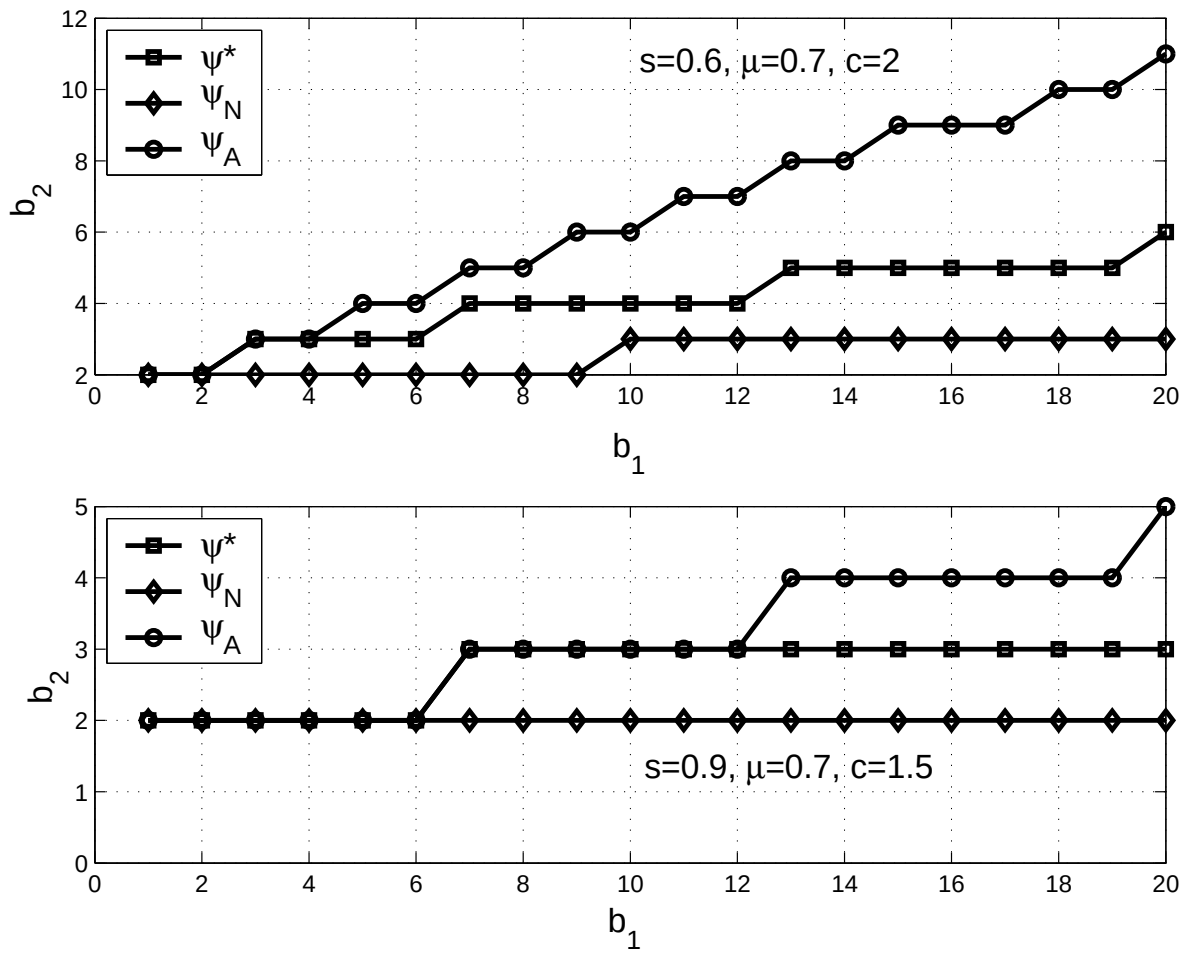

Fig. 4. Numerical Example 2

*Remark 7 (RAND and RFON have low computational complexity):* Note that the switchover curves $\psi_A$ and $\psi_N$, which bound the optimal switchover curve $\psi^{\star}$, can be computed in closed form as functions of the system parameters $s$, $\mu$, and $c$ by following the analysis in Section V-C. Consequently, the decision of the RAND policy can be obtained without explicitly solving any DP equations. Also, as numerical example 2 demonstrated, RAND provides a fairly good approximation to the optimal policy $\pi^{\star}$. Now recall that the decision of $\pi^{\star}$ is used in every time slot by the FON algorithm in a quasi-static fashion. Similarly, RFON uses the decision of RAND in every time slot. However, RFON has substantially lower implementation complexity in contrast to FON because, unlike FON, it does not require the solution to a set of DP equations in every time slot. We will compare the performance of FON and RFON in dynamic scenarios (varying $s$ and $\mu$) via simulations in Section VI.

## VI. Performance Evaluation

In this section, we evaluate the efficacy of the proposed fetching algorithms via simulations, and contrast it to benchmark algorithms. In particular, we simulate the following algorithms:

- *OPT*: The optimal fetching algorithm, computed by numerically solving the DP equations in (1). OPT is implemented via a lookup table, which is computed offline.
- *FON*: A quasi-static algorithm, which makes a fetching decision based on a numerical solution to the DP equations for the reduced model, as given by (2).
- *RFON*: Another quasi-static algorithm. RFON analytically bounds the optimal switchover curve for the reduced model (as computed by FON) and then constructs a randomized interpolation between the bounding curves to arrive at a fetching decision.
- *Always Fetch*: A benchmark algorithm, which always chooses to fetch a packet from the CS, as long as the queue at the CS is non-empty.
- *Never Fetch*: Another benchmark algorithm, which fetches a packet from the CS only when the queue at the MT is empty.

Recall from our discussion in Section II-C that the choice of per packet buffering cost rate at the MT, denoted $c$, determines the *congestion vs. latency tradeoff*, and hence, the operating point for the system. We will use this tradeoff curve as a *performance metric* to contrast the performance of different algorithms. Each point on the tradeoff curve is described a two-tuple: $(b_2^{\text{ave}}, d^{\text{ave}})$. Here $b_2^{\text{ave}}$ is the average backlog of $\mathcal{Q}_2$ (the queue at the MT) and $d^{\text{ave}}$ is the average end-to-end delay per task[§]. Both metrics are computed by averaging over an entire simulation run. Note that ${b_2}^{\text{ave}}$ is a measure of *congestion* at the MT, while $d^{\text{ave}}$ is a measure of the overall *latency* experienced by a typical task in the system. As discussed earlier, a congestion vs. latency tradeoff curve can be generated by varying the per packet per unit time cost parameter $c$. Given this performance metric, a policy $\pi$ is better than another policy $\pi'$ if $\pi$ yields a lower average backlog $b_2^{\text{ave}}$ than $\pi'$ for a fixed average delay $d^{\text{ave}}$, or a lower $d^{\text{ave}}$ for a fixed $b_2^{\text{ave}}$. Note that the *always fetch* and *never fetch* policies are oblivious to the cost parameter $c$, therefore for these policies, the entire tradeoff curve collapses to a single point.

We evaluate the performance of the five algorithms listed above under a wide variety of operational regimes. For all scenarios considered here, the probability of successful transmission over the wireless channel from the CS to the MT and average task execution time at the MT are modeled as two state Markov chains (as described in Section II). The state transition matrix for the wireless channel state is denoted:

$$P = \begin{bmatrix} p_{11} & 1 - p_{11} \\ 1 - p_{22} & p_{22} \end{bmatrix} \tag{8}$$

[§]The end-to-end delay for a task is comprised of four components: the waiting time in the queue at the Cs, transmission time over the wireless channel from the CS to the MT, waiting time in the queue at the MT, and processing time at the MT once the processor at the MT is allocated to the task.

and the state transition matrix for the processor state is denoted:

$$Q = \begin{bmatrix} q_{11} & 1 - q_{11} \\ 1 - q_{22} & q_{22} \end{bmatrix}. \tag{9}$$

The success probabilities in the two possible channel states are denoted by $s_1$ and $s_2$, and the average task execution times in the two possible processor states are denoted by $1/\mu_1$ and $1/\mu_2$, respectively (equivalently, the task execution rates are $\mu_1$ and $\mu_2$). Without loss of generality, we assume $s_1 < s_2$ and $\mu_1 > \mu_2$. Thus, the first channel state corresponds to a “good” channel and the second channel state corresponds to a “bad” channel. Similarly, the first processor state is one of “low” utilization, whereas the second processor state is one of “high” utilization. The probabilities $p_{11}$ and $p_{22}$ determine the frequency at which the wireless channel switches between its “good” and “bad” states. Similarly, the probabilities $q_{11}$ and $q_{22}$ determine the rate at which the processor at the MT switches between states of “low” and “high” utilization. We define two more derived parameters: $\delta_s \triangleq s_2 - s_1$ and $\delta_\mu \triangleq 1/\mu_2 - 1/\mu_1$. A large $\delta_s$ implies that the channel can enter deep fades relative to its “good” state. A large $\delta_\mu$ implies that average task processing times go up significantly (relative to the “low” utilization state) when the processor enters a “high” utilization state.

A variety of operational regimes can be envisioned, depending upon the values assumed by the parameters $p_{11}$, $p_{12}$, $q_{11}$, $q_{12}$, $\delta_s$, and $\delta_\mu$. we demonstrate via simulations that the proposed algorithms FON and RFON yield tradeoff curves comparable to the optimal tradeoff curve generated by OPT, under a wide range of operating scenarios. For all simulations, we assume that $\mathcal{Q}_1$, the queue at the CS, has 20 tasks initially, and no new tasks arrive to this queue over a simulation run. We vary the cost parameter $c$ from 1 to 100 to sweep the congestion vs. latency tradeoff curve.

### *A. Slow fading*

For the slow fading scenario, we assume $p_{11} = p_{22} = p = 0.9$. Thus, the expected sojourn time of the wireless channel in each state is $1/(1-p) = 10$ time slots. This essentially implies that each packet transmitted by the CS is likely to experience a static channel over all transmission attempts. This could well be the case in a mobile computing system deployed indoors, with static MTs and limited co-channel interference from other wireless networks. We fix $q_{11} = 0.5$, $q_{22} = 0.3$. We assume that the two tuple $(s_1, s_2)$ can take one of two possible values: $(0.1, 0.9)$ or $(0.4, 0.5)$. For the former case, $\delta_s = 0.8$, i.e., the channel can enter a deep fade relative to its “good” state. For the latter case, $\delta_s = 0.1$, i.e., the channel is fairly static over time. Further, we assume that the two tuple $(\mu_1, \mu_2)$ can take one of two possible values: $(0.9, 0.1)$ or $(0.6, 0.3)$. For the former case, $\delta_\mu = 8.89$, which implies that average task execution times vary significantly from one processor state to the other. For the latter case, $\delta_\mu = 1.67$, which means that average task execution times are fairly constant over time. Thus, we have the following four subcases:

($\delta_s = 0.8, \delta_\mu = 8.89$), ($\delta_s = 0.1, \delta_\mu = 1.67$), ($\delta_s = 0.1, \delta_\mu = 8.89$), and ($\delta_s = 0.8, \delta_\mu = 1.67$).

The congestion vs. latency tradeoff curves for the four cases are depicted in Figs. 5, 6, 7, and 8, respectively.

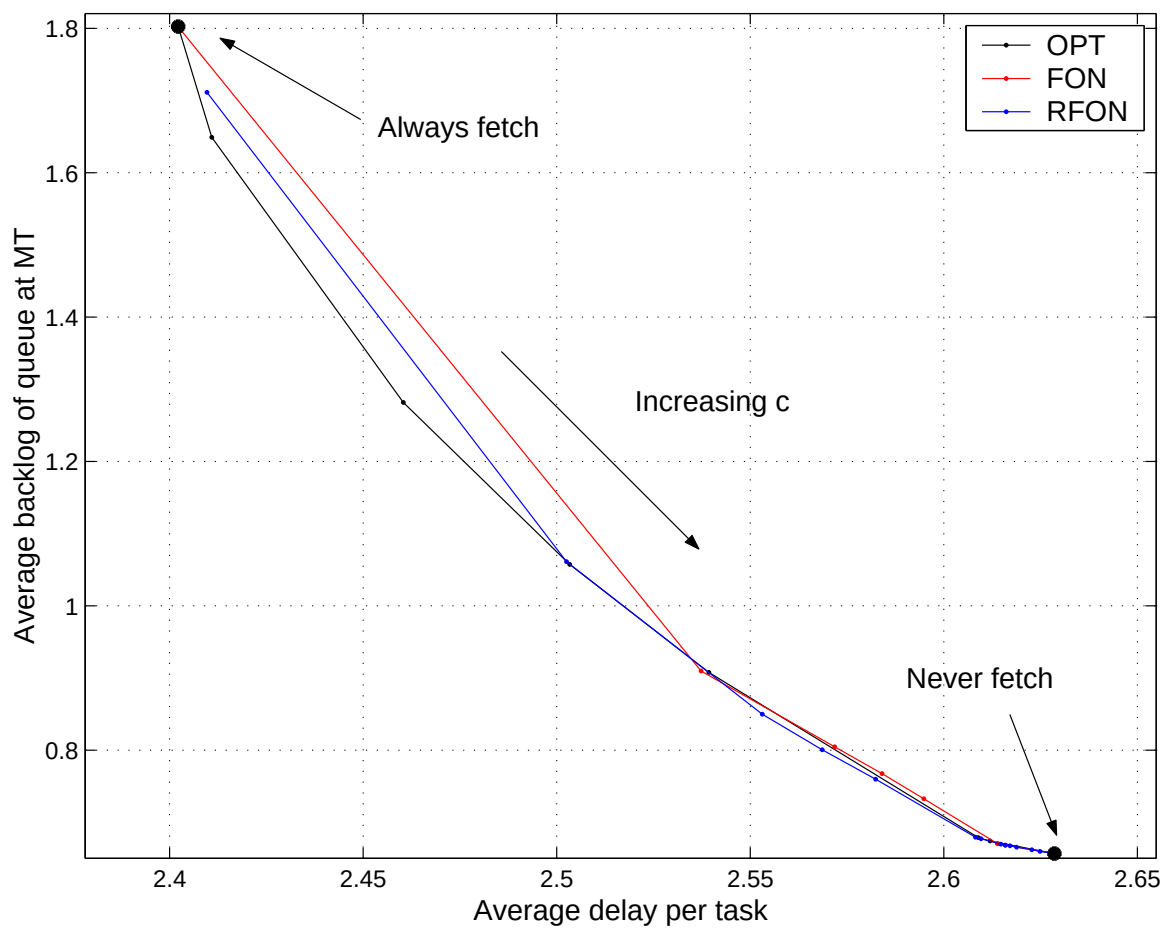

Fig. 5. Slow fading ($p_{11} = p_{22} = 0.9$), $q_{11} = 0.5$, $q_{22} = 0.3$, $\delta_s = 0.8$, and $\delta_\mu = 8.89$.

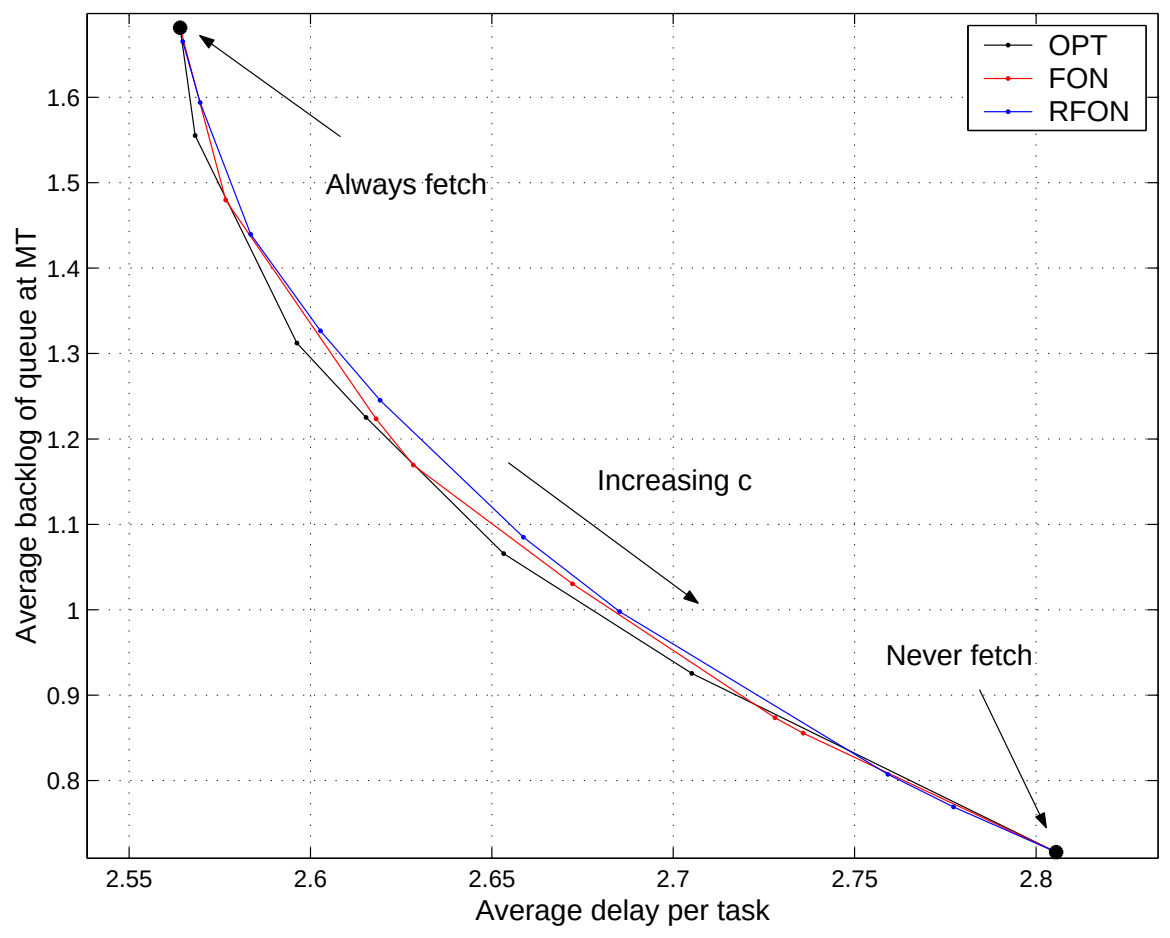

Fig. 6. Slow fading ($p_{11} = p_{22} = 0.9$), $q_{11} = 0.5$, $q_{22} = 0.3$, $\delta_s = 0.1$, and $\delta_\mu = 1.67$.

## B. Fast fading

For the fast fading scenario, we assume $p_{11} = p_{22} = p = 0.1$. Thus, the expected sojourn time of the wireless channel in each state is $1/(1-p) = 1.1$ time slots. The implication is that the channel state changes almost every time slot, so a task will experience a different channel state for every retransmission. This could be the case if the MT is mobile or operates in an environment where co-channel interference fluctuates rapidly relative to the dynamics of the system. Similar to the slow fading case, we consider four subcases, depending upon the magnitude of $\delta_s$ and $\delta_\mu$. As before, these four subcases are given by: ($\delta_s = 0.8, \delta_\mu = 8.89$), ($\delta_s = 0.1, \delta_\mu = 1.67$), ($\delta_s = 0.1, \delta_\mu = 8.89$), and ($\delta_s = 0.8, \delta_\mu = 1.67$). For all subcases, we fix $q_{11} = 0.5$, $q_{22} = 0.3$.

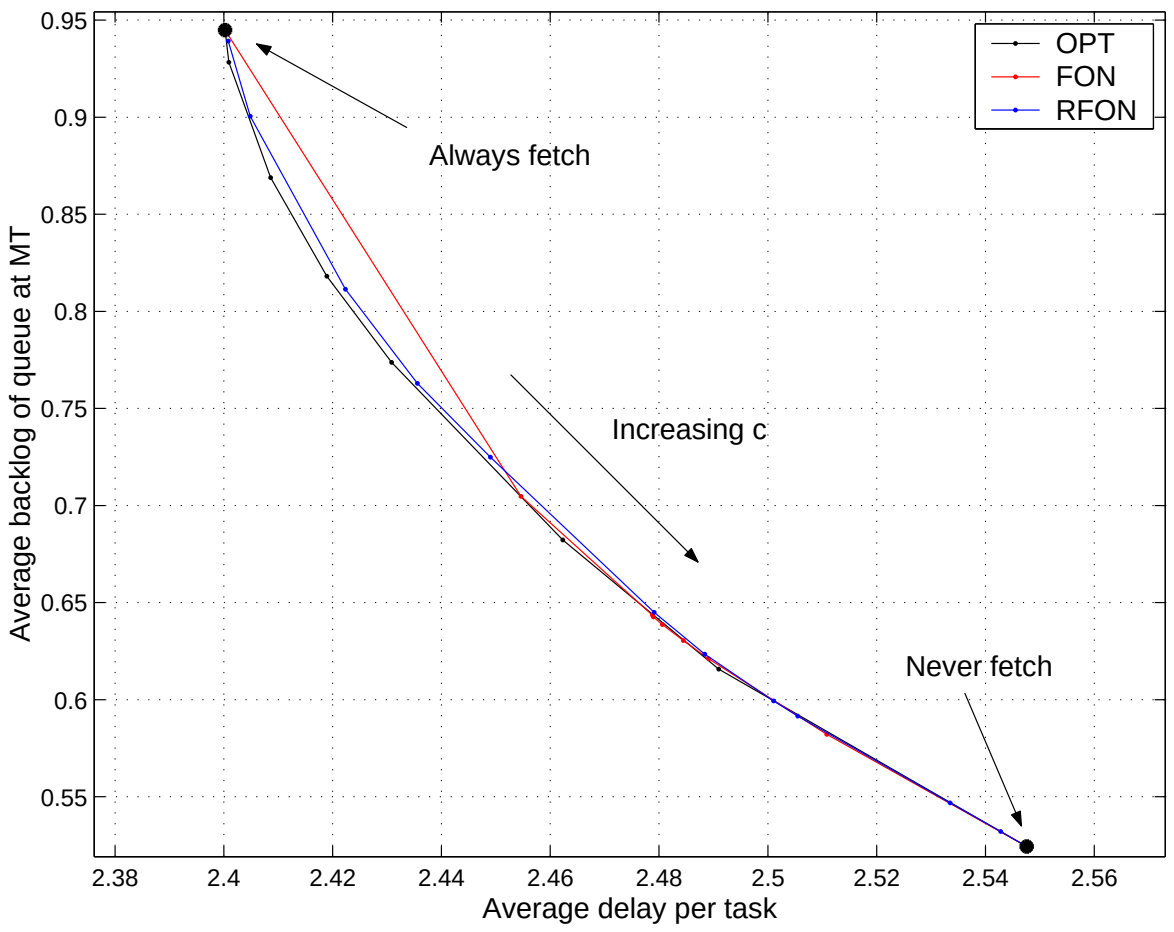

Fig. 7. Slow fading ($p_{11} = p_{22} = 0.9$), $q_{11} = 0.5$, $q_{22} = 0.3$, $\delta_s = 0.1$, and $\delta_\mu = 8.89$.

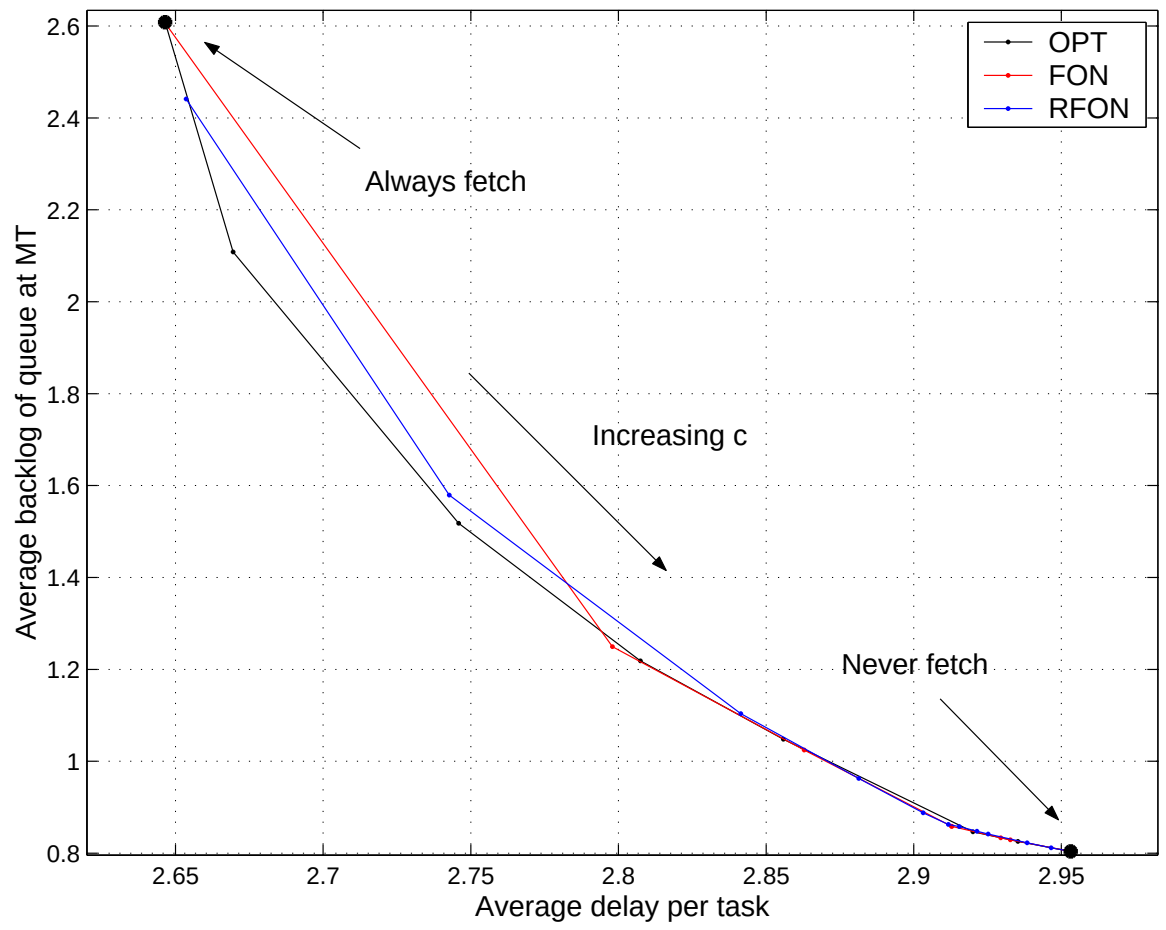

Fig. 8. Slow fading ($p_{11} = p_{22} = 0.9$), $q_{11} = 0.5$, $q_{22} = 0.3$, $\delta_s = 0.8$, and $\delta_\mu = 1.67$.

The congestion vs. latency tradeoff curves for the four cases are depicted in Figs. 9, 10, 11, and 12, respectively.

For all eight operational regimes considered here, we observe that the tradeoff curves generated by FON and RFON are very similar to the tradeoff curve generated by OPT. Note that the optimal policy OPT is assumed to have a priori knowledge of all system parameters, viz., $P$, $Q$, $(s_1, s_2)$, and $(\mu_1, \mu_2)$. In contrast, FON and RFON only assumed to know the instantaneous values of $s$ and $\mu$. These policies have no knowledge of the possible values either $s$ or $\mu$ can assume, or the underlying statistics (Markovian in our examples) which govern $s$ and $\mu$. The decisions of both FON and RFON are based entirely on the instantaneous operating point of the system through a quasi-static approximation, which assumes that the current operating point will persist forever in the future. All three policies have access to the backlogs of both $\mathcal{Q}_1$ and $\mathcal{Q}_2$.

As expected, OPT offers the best tradeoff in all operating scenarios. FON and RFON match the performance of OPT quite closely in slow fading conditions, and also when $\delta_s$ and $\delta_\mu$ are small. This is because the quasi-static approximation is quite accurate in conditions which fluctuate slowly, and do not vary much whenever they fluctuate. Consistent with this intuition, the biggest departure of FON and RFON from OPT is observed in fast fading, when $\delta_s$ and/or $\delta_\mu$ is large.

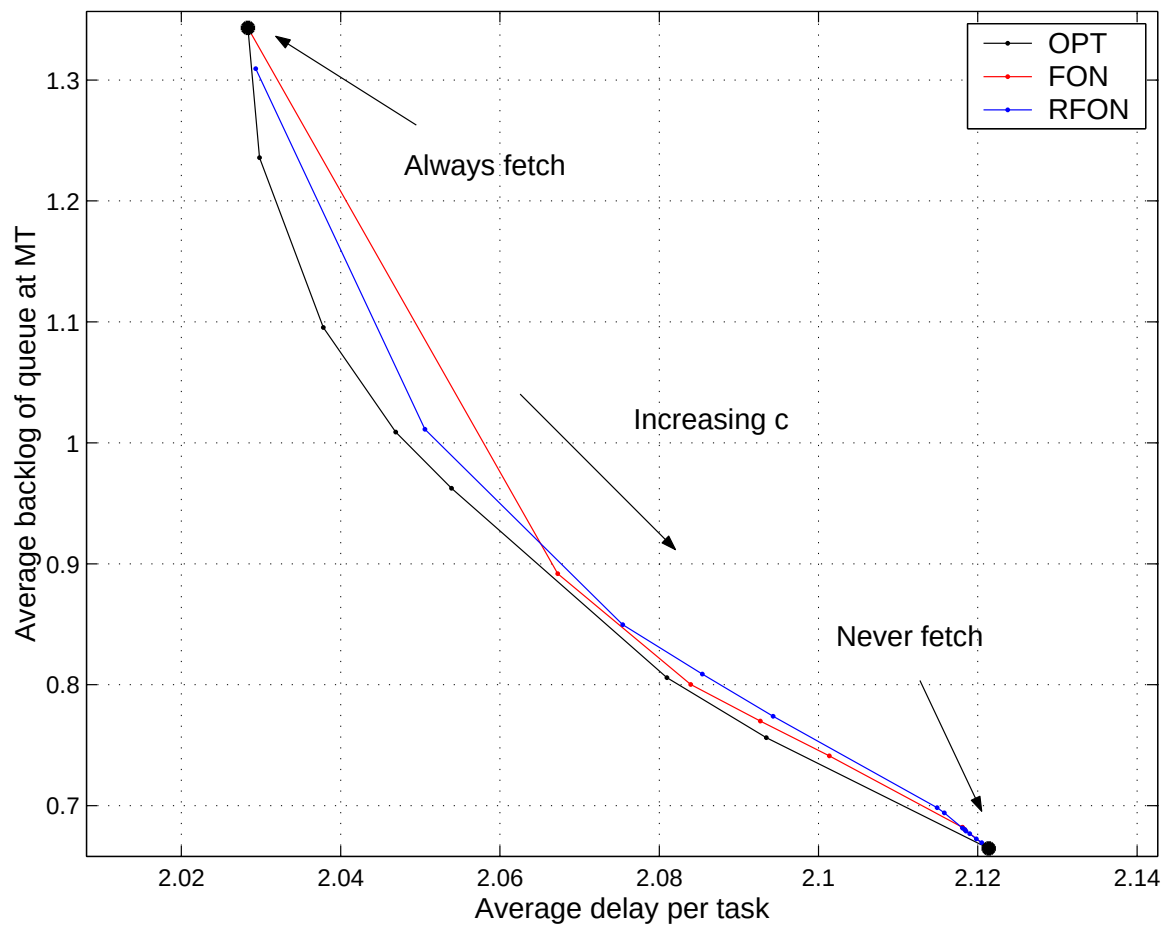

Fig. 9. Fast fading ($p_{11} = p_{22} = 0.1$), $q_{11} = 0.5$, $q_{22} = 0.3$, $\delta_s = 0.8$, and $\delta_\mu = 8.89$.

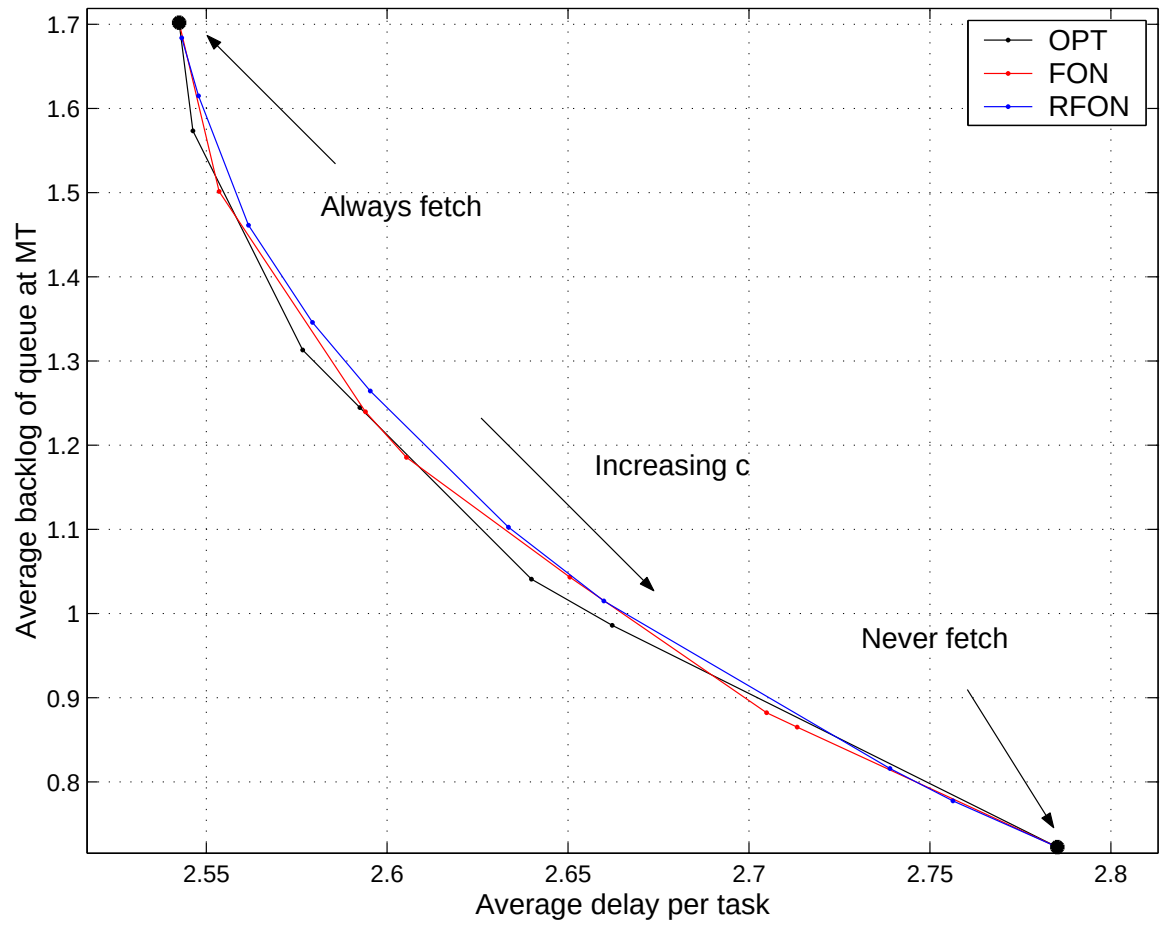

Fig. 10. Fast fading ($p_{11} = p_{22} = 0.1$), $q_{11} = 0.5$, $q_{22} = 0.3$, $\delta_s = 0.1$, and $\delta_\mu = 1.67$.

Finally, note that the always fetch and never fetch policies appear as two extreme points on the optimal tradeoff curve. Thus, the always fetch policy seems to be optimal in the extreme regime where a large average congestion $b_2^{\text{ave}}$ can be tolerated for a small latency $d^{\text{ave}}$. Similarly, the never fetch policy appears to be the best choice in the extreme regime where a small $b_2^{\text{ave}}$ is desired, even at the expense of a large $d^{\text{ave}}$. However, none of these policies have the ability to provide a tradeoff between $b_2^{\text{ave}}$ and $d^{\text{ave}}$. Also, it is important to note that the optimal tradeoff curve, as well as the tradeoff curves for FON and RFON are convex, which means that the tradeoff curve generated by time sharing (either randomly or deterministically) between the simplistic always fetch and never fetch policies will be strictly worse than the proposed policies. This argument justifies the small increase in complexity afforded by FON and RFON (relative to the simplistic always fetch and never fetch) in order to enhance system performance.

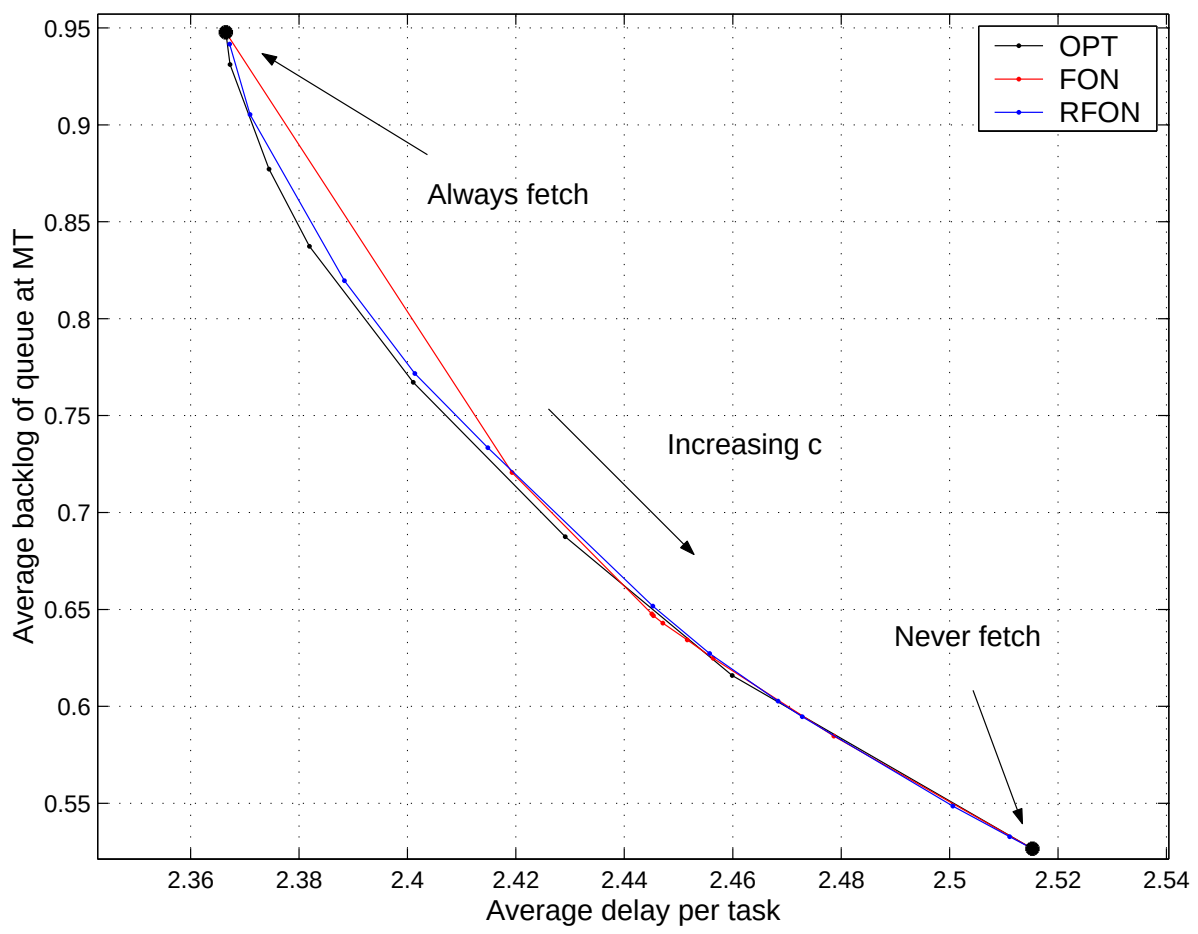

Fig. 11. Fast fading ($p_{11} = p_{22} = 0.1$), $q_{11} = 0.5$, $q_{22} = 0.3$, $\delta_s = 0.1$, and $\delta_\mu = 8.89$.

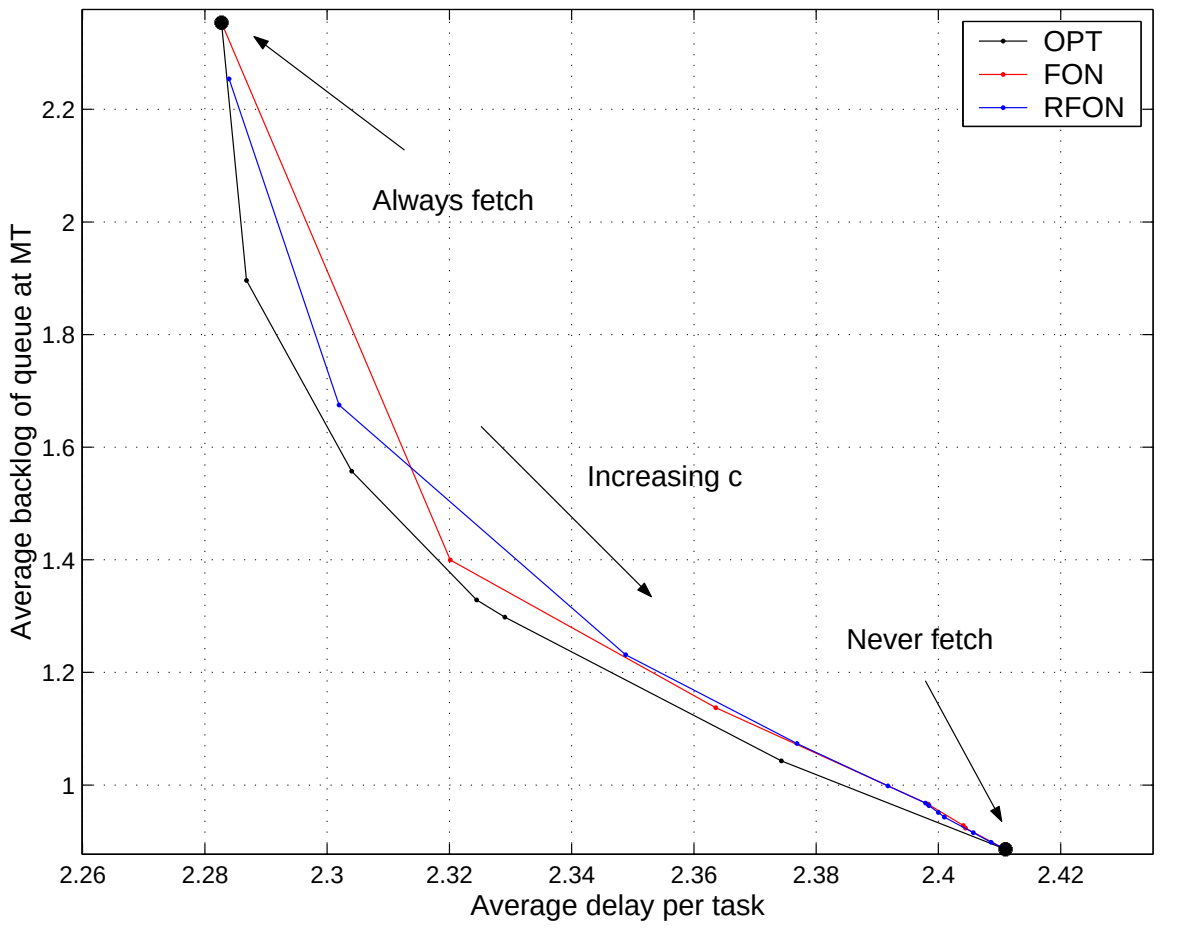

Fig. 12. Fast fading ($p_{11} = p_{22} = 0.1$), $q_{11} = 0.5$, $q_{22} = 0.3$, $\delta_s = 0.8$, and $\delta_\mu = 1.67$.

## VII. Conclusions

A gamut of mobile application pose diverse computation and processing requirements for battery power and memory constrained mobile devices. Judicious prefetching of computation tasks at the mobile terminals is thus important. Aggressive prefetching can result in congestion at the mobile device while lazy retrieval of these tasks can cause increase in latency thereby resulting in degraded application performance. We examine this buffering versus latency trade-off under varying wireless channel conditions and mobile terminal congestion states via dynamic programming methodology. We suggest quasi-static and randomized algorithms that alleviate the computational complexity of the dynamic programming solutions. Through evaluation experiments under slow and fast channel conditions we should that our low complexity heuristic prefetching algorithms come quite close in performance to the optimal solution.

## VIII. Appendix

### A. Proof of Theorem 1

We use value iteration and the principle of mathematical induction to prove the theorem. The DP equations in (2) can be solved by using the method of backward value iteration, where the estimate of the value function in state $\mathbf{b}$ at time $n$ iteration, namely $V^n(\mathbf{b})$, is expressed in terms of the estimate of value function at time $(n+1)$, namely $V^{n+1}(\cdot)$ as:

$$\begin{aligned} V^n(\mathbf{b}) = \min\{&\mu V^{n+1}(\mathbf{b}-\mathbf{e}_2) + \bar{\mu}V^{n+1}(\mathbf{b}), s\mu V^{n+1}(\mathbf{b}-\mathbf{e}_1) + \bar{s}\mu V^{n+1}(\mathbf{b}-\mathbf{e}_2) \\ &+ s\bar{\mu}V^{n+1}(\mathbf{b}-\mathbf{e}_1+\mathbf{e}_2) + \bar{s}\bar{\mu}V^{n+1}(\mathbf{b})\} + \langle \mathbf{c}, \mathbf{b}\rangle, \end{aligned} \tag{10}$$

along with the boundary conditions $V^{N+1}(\mathbf{b}) = 0 \; \forall \, \mathbf{b}$, where $N$ is the length of the horizon over which the value iteration equations in (10) are solved. It is known that under certain conditions, $V^0(\mathbf{b})$ converges to the true value function $V(\mathbf{b})$, which satisfies (2), as $N \to \infty$. For guaranteed convergence, it is sufficient that $\exists \, M \in \mathbb{N}$, such that the system reaches the terminal state $(0, 0)$ under any admissible policy within $M$ steps, regardless of the initial state. This condition is clearly satisfied by our two queue tandem as long as $s, \mu > 0$.

Now, to establish the desired result, we will show that for any $N$, if the switchover property is satisfied at time $(n+1)$, then it is also satisfied at time $n$. Inductively, this implies that the switchover property is satisfied at time $0 \; \forall \, N$, and thus satisfied by the optimal policy $\pi^\star$ in the limit $N \to \infty$.

Define,

$$\omega^n(\mathbf{b}) \triangleq s\mu[V^{n+1}(\mathbf{b} - \mathbf{e}_2) - V^{n+1}(\mathbf{b} - \mathbf{e}_1)] + s\bar{\mu}[V^{n+1}(\mathbf{b}) - V^{n+1}(\mathbf{b} - \mathbf{e}_1 + \mathbf{e}_2)]. \tag{11}$$

It is easy to verify that the optimal decision at time $n$ in state $\mathbf{b}$ is $\overline{\text{FE}}$ if $\omega^n(\mathbf{b}) \leq 0$, and FE otherwise.

We now fix $N$ and assume that the optimal policy at time $n + 1 < N$, denoted $\pi^\star_{n+1,N}$ satisfies the switchover property. From the definition of $\omega^{n+1}(\mathbf{b})$, it follows that the switchover property can equivalently be interpreted as follows: $\omega^{n+1}(b_1, b_2)$ is a non-decreasing function of $b_1$ and a non-increasing function of $b_2$. Based on this interpretation, we want to show that $\omega^n(\mathbf{b})$ is a non-decreasing function of $b_1$ and a non-increasing function of $b_2$, i.e., $\pi^\star_{n,N}$ also satisfies the switchover property.

Note that the optimality of the switchover property at time $(n+1)$ implies that the optimal policy $\pi^\star_{n+1,N}$ splits the state space, i.e., the $(b_1, b_2)$ plane into two distinct *decision regions*, corresponding to the two possible decisions FE and $\overline{\text{FE}}$, respectively.

Now, it follows from (11) that $\omega^n(\mathbf{b})$ is a function of $V^{n+1}(\mathbf{b} - \mathbf{e}_1)$, $V^{n+1}(\mathbf{b} - \mathbf{e}_2)$, $V^{n+1}(\mathbf{b})$, and $V^{n+1}(\mathbf{b} - \mathbf{e}_1 + \mathbf{e}_2)$. By the same token, $\omega^n(\mathbf{b} + \mathbf{e}_1)$ is a function of $V^{n+1}(\mathbf{b})$, $V^{n+1}(\mathbf{b} + \mathbf{e}_1 - \mathbf{e}_2)$, $V^{n+1}(\mathbf{b} + \mathbf{e}_1)$, and $V^{n+1}(\mathbf{b} + \mathbf{e}_2)$. Thus, to show that $\omega^n(\mathbf{b} + \mathbf{e}_1) \geq \omega^n(\mathbf{b})$, we need to consider numerous cases, depending on the optimal decision at time $(n + 1)$ in the states of interest listed above. In the interest of space, we are only going to consider two representative cases here: (i) all states of interest lie in the decision region corresponding to action $\overline{\text{FE}}$, and (ii) all states of interest lie in the decision region corresponding to action FE. All other cases where the boundary between the decision regions splits the states of interest into two sets can be treated as a combination of the two representative cases. Also, we will only focus on establishing the monotonicity of $\omega^n(b_1, b_2)$ in its first argument. The proof of monotonicity of $\omega^n(b_1, b_2)$ in its second argument follows analogously.

*Case 1:* We assume that the optimal decision at time $(n + 1)$ is $\overline{\text{FE}}$ in the following set of states: $\mathcal{X} = \{\mathbf{b}, \mathbf{b} \pm \mathbf{e}_1, \mathbf{b} \pm \mathbf{e}_2, \mathbf{b} + \mathbf{e}_1 - \mathbf{e}_2, \mathbf{b} - \mathbf{e}_1 + \mathbf{e}_2\}$. Thus for any state $\mathbf{x} \in \mathcal{X}$, we have

$$V^{n+1}(\mathbf{x}) = \mu V^{n+1}(\mathbf{x} - \mathbf{e}_2) + \bar{\mu} V^{n+1}(\mathbf{x}) + \langle \mathbf{c}, \mathbf{x} \rangle \quad \forall \, \mathbf{x} \in \mathcal{X}. \tag{12}$$

For convenience, define

$$\Delta^n(\mathbf{b}) \triangleq V^n(\mathbf{b}-\mathbf{e}_2) - V^n(\mathbf{b}-\mathbf{e}_1). \tag{13}$$

If follows from (11) and (13) that

$$\omega^n(\mathbf{b}) = s\mu\Delta^{n+1}(\mathbf{b}) + s\bar{\mu}\Delta^{n+1}(\mathbf{b}+\mathbf{e}_2). \tag{14}$$

From (12) and (13) it follows:

$$\begin{aligned} \Delta^{n+1}(\mathbf{b}) &= \mu\Delta^{n+2}(\mathbf{b}-\mathbf{e}_2) + \bar{\mu}\Delta^{n+2}(\mathbf{b}) + \langle \mathbf{c}, \mathbf{e}_1 - \mathbf{e}_2 \rangle \\ \Delta^{n+1}(\mathbf{b}+\mathbf{e}_2) &= \mu\Delta^{n+2}(\mathbf{b}) + \bar{\mu}\Delta^{n+2}(\mathbf{b}+\mathbf{e}_2) + \langle \mathbf{c}, \mathbf{e}_1 - \mathbf{e}_2 \rangle \end{aligned} \tag{15}$$

Substituting (15) in (14) yields

$$\begin{aligned} \omega^n(\mathbf{b}) = \mu[&\underbrace{s\mu\Delta^{n+2}(\mathbf{b}-\mathbf{e}_2) + s\bar{\mu}\Delta^{n+2}(\mathbf{b})}_{=\omega^{n+1}(\mathbf{b}-\mathbf{e}_2) \text{ from } (14)}] \\ &+ \bar{\mu}[\underbrace{s\mu\Delta^{n+2}(\mathbf{b}) + s\bar{\mu}\Delta^{n+2}(\mathbf{b}+\mathbf{e}_2)}_{=\omega^{n+1}(\mathbf{b}) \text{ from } (14)}] + s\langle \mathbf{c}, \mathbf{e}_1 - \mathbf{e}_2 \rangle, \end{aligned} \tag{16}$$

implying $\omega^n(\mathbf{b}) = \mu\omega^{n+1}(\mathbf{b}-\mathbf{e}_2) + \bar{\mu}\omega^{n+1}(\mathbf{b}) + s(1-c)$. By the same token, $\omega^n(\mathbf{b}+\mathbf{e}_1) = \mu\omega^{n+1}(\mathbf{b}+\mathbf{e}_1-\mathbf{e}_2) + \bar{\mu}\omega^{n+1}(\mathbf{b}+\mathbf{e}_1) + s(1-c)$. It now easily follows from our inductive assumption ($\omega^{n+1}(b_1, b_2)$ is a non-decreasing function of $b_1$) that $\omega^n(\mathbf{b}+\mathbf{e}_1) \geq \omega^n(\mathbf{b})$, as desired.

*Case 2:* We now assume that the optimal decision in all states in the set $\mathcal{X}$ (as defined for Case 1) at time $(n+1)$ is FE. Thus, for any $\mathbf{x} \in \mathcal{X}$ we have

$$V^{n+1}(\mathbf{x}) = s\mu V^{n+2}(\mathbf{x}-\mathbf{e}_1) + \bar{s}\mu V^{n+2}(\mathbf{x}-\mathbf{e}_2) + s\bar{\mu}V^{n+2}(\mathbf{x}-\mathbf{e}_1+\mathbf{e}_2) + \bar{s}\bar{\mu}V^{n+2}(\mathbf{x}) + \langle \mathbf{c}, \mathbf{x} \rangle. \tag{17}$$

Following the definition of $\Delta^n(\mathbf{b})$ in (13), we have

$$\Delta^{n+1}(\mathbf{b}) = s\mu\Delta^{n+2}(\mathbf{b}-\mathbf{e}_1) + \bar{s}\mu\Delta^{n+2}(\mathbf{b}-\mathbf{e}_2) + s\bar{\mu}\Delta^{n+2}(\mathbf{b}-\mathbf{e}_1+\mathbf{e}_2) + \bar{s}\bar{\mu}\Delta^{n+2}(\mathbf{b}) + \langle \mathbf{c}, \mathbf{e}_1 - \mathbf{e}_2 \rangle. \tag{18}$$

$$\Delta^{n+1}(\mathbf{b}+\mathbf{e}_2) = s\mu\Delta^{n+2}(\mathbf{b}-\mathbf{e}_1+\mathbf{e}_2) + \bar{s}\mu\Delta^{n+2}(\mathbf{b}) + s\bar{\mu}\Delta^{n+2}(\mathbf{b}-\mathbf{e}_1+2\mathbf{e}_2) + \bar{s}\bar{\mu}\Delta^{n+2}(\mathbf{b}+\mathbf{e}_2) + \langle \mathbf{c}, \mathbf{e}_1 - \mathbf{e}_2 \rangle. \tag{19}$$

Substituting (18) and (19) in (14) and re-arrangement of terms yields

$$\omega^n(\mathbf{b}) = s\mu\omega^{n+1}(\mathbf{b}-\mathbf{e}_1) + \bar{s}\mu\omega^{n+1}(\mathbf{b}-\mathbf{e}_2) + s\bar{\mu}\omega^{n+1}(\mathbf{b}-\mathbf{e}_1+\mathbf{e}_2) + \bar{s}\bar{\mu}\omega^{n+1}(\mathbf{b}) + s\langle \mathbf{c}, \mathbf{e}_1 - \mathbf{e}_2 \rangle. \tag{20}$$

By the same token,

$$\omega^n(\mathbf{b}+\mathbf{e}_1) = s\mu\omega^{n+1}(\mathbf{b}) + \bar{s}\mu\omega^{n+1}(\mathbf{b}+\mathbf{e}_1-\mathbf{e}_2) + s\bar{\mu}\omega^{n+1}(\mathbf{b}+\mathbf{e}_2) + \bar{s}\bar{\mu}\omega^{n+1}(\mathbf{b}+\mathbf{e}_1) + s\langle \mathbf{c}, \mathbf{e}_1 - \mathbf{e}_2 \rangle. \tag{21}$$

It now easily follows from (20), (21), and our inductive assumption ($\omega^{n+1}(b_1, b_2)$ is a non-decreasing function of $b_1$) that $\omega^n(\mathbf{b} + \mathbf{e}_1) \geq \omega^n(\mathbf{b})$, as desired.

All other cases not covered by case 1 and case 2 above are of the type: $\mathcal{X} = \mathcal{X}_{\text{FE}} \cup \mathcal{X}_{\overline{\text{FE}}}, \mathcal{X}_{\text{FE}} \cap \mathcal{X}_{\overline{\text{FE}}}$, where the optimal decision at time $(n+1)$ in state $\mathbf{x} \in \mathcal{X}$ is FE if $\mathbf{x} \in \mathcal{X}_{\text{FE}}$, and $\overline{\text{FE}}$ if $\mathbf{x} \in \mathcal{X}_{\overline{\text{FE}}}$. These cases can be treated as a "combination" of two extreme cases $\mathcal{X}_{\text{FE}} = \emptyset$ (Case 1) and $\mathcal{X}_{\overline{\text{FE}}} = \emptyset$ (Case 2) treated above. We skip the details for brevity. Finally, the proof for $\omega^n(\mathbf{b} + \mathbf{e}_2) \leq \omega^n(\mathbf{b})$ is analogous to the proof above.

It thus follows from the principle of mathematical induction that for any $N \in \mathbb{N}$, $\omega^n(b_1, b_2)$ is a non-decreasing function of $b_1$ and a non-increasing function of $b_2$ $\forall\ n = 0, \ldots, N$. The equivalence between the monotonicity of $\omega^n(\mathbf{b})$ and the optimality of a switchover policy implies that the optimal policy at each time $n$, namely $\pi^{\star}_{n,N}$, satisfies the switchover property, for any fixed $N$. Finally, it follows from the convergence of the value iteration algorithm that the optimal policy $\pi^{\star}$, which satisfies the Bellman's equations in (2) is of switchover type, as claimed.

### B. *Proof of Lemma 1*

From the system dynamics described in Section II-D, it follows that

$$C_N(\mathbf{b}) = \begin{cases} \mu C_N(\mathbf{b} - \mathbf{e}_2) + \bar{\mu} C_N(\mathbf{b}) + \langle \mathbf{c}, \mathbf{b} \rangle & ; \quad b_2 > 0 \\ s\mu C_N(\mathbf{b} - \mathbf{e}_1) + s\bar{\mu} C_N(\mathbf{b} - \mathbf{e}_1 + \mathbf{e}_2) + \bar{s} C_N(\mathbf{b}) + \langle \mathbf{c}, \mathbf{b} \rangle & ; \quad b_2 = 0. \end{cases}$$

Rearranging and combining terms we get

$$C_N(\mathbf{b}) = \begin{cases} C_N(\mathbf{b} - \mathbf{e}_2) + \dfrac{\langle \mathbf{c}, \mathbf{b} \rangle}{\mu} & ; \quad b_2 > 0 \\ C_N(b_1 - 1, 0) + \left( \dfrac{\bar{\mu}}{\mu} + \dfrac{1}{s} \right) b_1 + \dfrac{(c-1)\bar{\mu}}{\mu} & ; \quad b_2 = 0, \end{cases}$$

implying $C_N(b_1, 0) = \left( \dfrac{\bar{\mu}}{\mu} + \dfrac{1}{s} \right) \dfrac{b_1^2}{2} + \left[ \dfrac{1}{2} \left( \dfrac{\bar{\mu}}{\mu} + \dfrac{1}{s} \right) + \dfrac{(c-1)\bar{\mu}}{\mu} \right] b_1$ and

$$C_N(\mathbf{b}) = C_N(b_1, 0) + \frac{b_1 b_2}{\mu} + \frac{c b_2 (b_2 + 1)}{2\mu},$$

which is the desired result.

### C. *Approximating $C_A(\mathbf{b})$ using a fluid caricature model*

*1) A fluid caricature model:* The fluid caricature model mimics the "mean behavior" of the time slotted, packet based model. The key attributes of the fluid caricature model are:

- $\mathcal{Q}_1$ and $\mathcal{Q}_2$ buffer infinitesimally divisible "fluids".
- Time is continuous.
- Fluid flows at a constant rate $s$ from $\mathcal{Q}_1$ to $\mathcal{Q}_2$ and flows out of $\mathcal{Q}_2$ at a constant rate $\mu$.

- A backlog cost at unit rate for every unit of fluid is incurred at $\mathcal{Q}_1$, and a congestion cost at a rate $c$ for every unit of fluid is incurred at $\mathcal{Q}_2$.

Similar to the time slotted model, no fluid arrives to $\mathcal{Q}_1$ after time $t = 0$.

*2) Proof of Lemma 2:* We need to consider two distinct cases:

1) $s \geq \mu$: Since $s \geq \mu$, $\mathcal{Q}_1$ drains faster than $\mathcal{Q}_2$. Denoting the initial amount of fluid in $\mathcal{Q}_1$ and $\mathcal{Q}_2$ by $b_1$ and $b_2$ respectively, $\mathcal{Q}_1$ first becomes empty at time $T_0 \triangleq b_1/s$ and stays empty thereafter. For $t \leq T_0$, the amount of fluid in $\mathcal{Q}_1$ at time $t$ is given by $b_1 - st$. Thus, the total backlog cost incurred at $\mathcal{Q}_1$ over the interval $[0, T_0]$ is $\int_0^{T_0} (b_1 - st)\ dt = \frac{b_1^2}{2s}$. Over the same interval, the amount of fluid in $\mathcal{Q}_2$ at time $t$ is given $b_2 + (s - \mu)t$. Thus, the total congestion cost incurred at $\mathcal{Q}_2$ over $[0, T_0]$ is $\int_0^{T_0} c(b_2 + (s - \mu)t)\ dt = \frac{cb_1b_2}{s} + \frac{cb_1^2}{2s}\left(1 - \frac{\mu}{s}\right)$. Now, the amount of fluid in $\mathcal{Q}_2$ at time $T_0$ is $B_2(T_0) = b_2 + b_1\left(1 - \frac{\mu}{s}\right)$. Thus, $\mathcal{Q}_2$ drains completely at time $T_0' = T_0 + \frac{B_2(T_0)}{\mu}$ and the amount of fluid in $\mathcal{Q}_2$ at time $t$ for $t \in [T_0, T_0']$ is given by $B_2(T_0) - \mu(t - T_0)$. Consequently, the congestion cost incurred over the interval $[T_0, T_0']$ at $\mathcal{Q}_2$ is given by $\int_{T_0}^{T_0'} c(B_2(T_0) - \mu(t - T_0))\ dt = \frac{c}{2\mu}\left[b_2 + b_1\left(1 - \frac{\mu}{s}\right)\right]^2$. The total cost $C_A^f(\mathbf{b})$ for the case $s \geq \mu$ is given by the sum of the three costs computed above.
2) $s < \mu$: We need to consider two further sub-cases. To this end, define $T_1 \triangleq b_2/(\mu - s)$ and let $T_0 = b_1/s$, as before. If $T_1 \leq T_0$, then $\mathcal{Q}_2$ drains before $\mathcal{Q}_1$. The backlog cost incurred at $\mathcal{Q}_1$ over the interval $[0, T_1]$ is $\int_0^{T_1} (b_1 - st)\ dt = b_1T_1 - \frac{sT_1^2}{2}$ and the corresponding congestion cost incurred at $\mathcal{Q}_2$ is $\int_0^{T_1} c(b_2 - (\mu - s)t)\ dt = c\left(b_2T_1 - \frac{(\mu - s)T_1^2}{2}\right)$. The amount of fluid in $\mathcal{Q}_1$ at the end of the interval is $B_1(T_1) = b_1 - sT_1$. An additional backlog cost of $\int_{T_1}^{T_0} (B_1(T - 1) - st)\ dt = B_1(T_1)(T_0 - T_1) - \frac{s(T_0 - T_1)^2}{2}$ is incurred over the interval $[T_1, T_0]$ at $\mathcal{Q}_1$. The cost incurred at $\mathcal{Q}_2$ over this interval is negligible. Thus, $C_A^f(\mathbf{b})$ for the case $s < \mu, T_1 \geq T_0$ is given by the sum of the three costs computed above.

   If $T_1 < T_0$, $\mathcal{Q}_2$ drains before $\mathcal{Q}_1$. The computation of $C_A^f$ in this case is identical to the case $s \geq \mu$ considered above.

*3) "Goodness" of approximation:* In this section, we explore the efficacy of $C_A^f(\mathbf{b})$ as an approximation to $C_A(\mathbf{b})$ via a numerical example.

*Numerical Example 3:* This example illustrates the goodness of $C_A^f(\mathbf{b})$ as an approximation to $C_A(\mathbf{b})$. The left side of Fig. 13 shows the fractional approximation error as a function of $b_2$ for three different values of $b_1$ for the case $s = 0.8$, $\mu = 0.6$ ($s > \mu$). The right side of the figure depicts the same plots for the case $s = 0.6$, $\mu = 0.8$ ($s < \mu$). Observe that the accuracy of the approximation increases as $b_1$ and $b_2$ increase. The relative error is below 5% for moderately large values of $b_1$ and $b_2$. The error is as much as 30% for small values of $b_1$ and $b_2$. For these cases, however, $C_A(\mathbf{b})$ can be computed exactly, with only a few computations. The "kinks" in the plot on the right correspond to the points at which $T_1 > T_0$ exceeds $T_0$ (as defined in Lemma 2). Note that $T_0$ is fixed on each curve, since $b_1$ is fixed on each curve. Further, $T_1$ increases linearly with $b_2$ on each curve. For fixed $b_1$, $T_1 > T_0$ implies $b_2 > (\mu/s - 1)b_1$, which is always satisfied for $\mu \leq s$ (left side of the figure). When $\mu > s$, with the values chosen in this example ($\mu = 0.8$, $s = 0.6$), the condition reduces to $b_2 > b_1/3$ (determining the points at which the kinks appear in the plot on the right side of the figure).

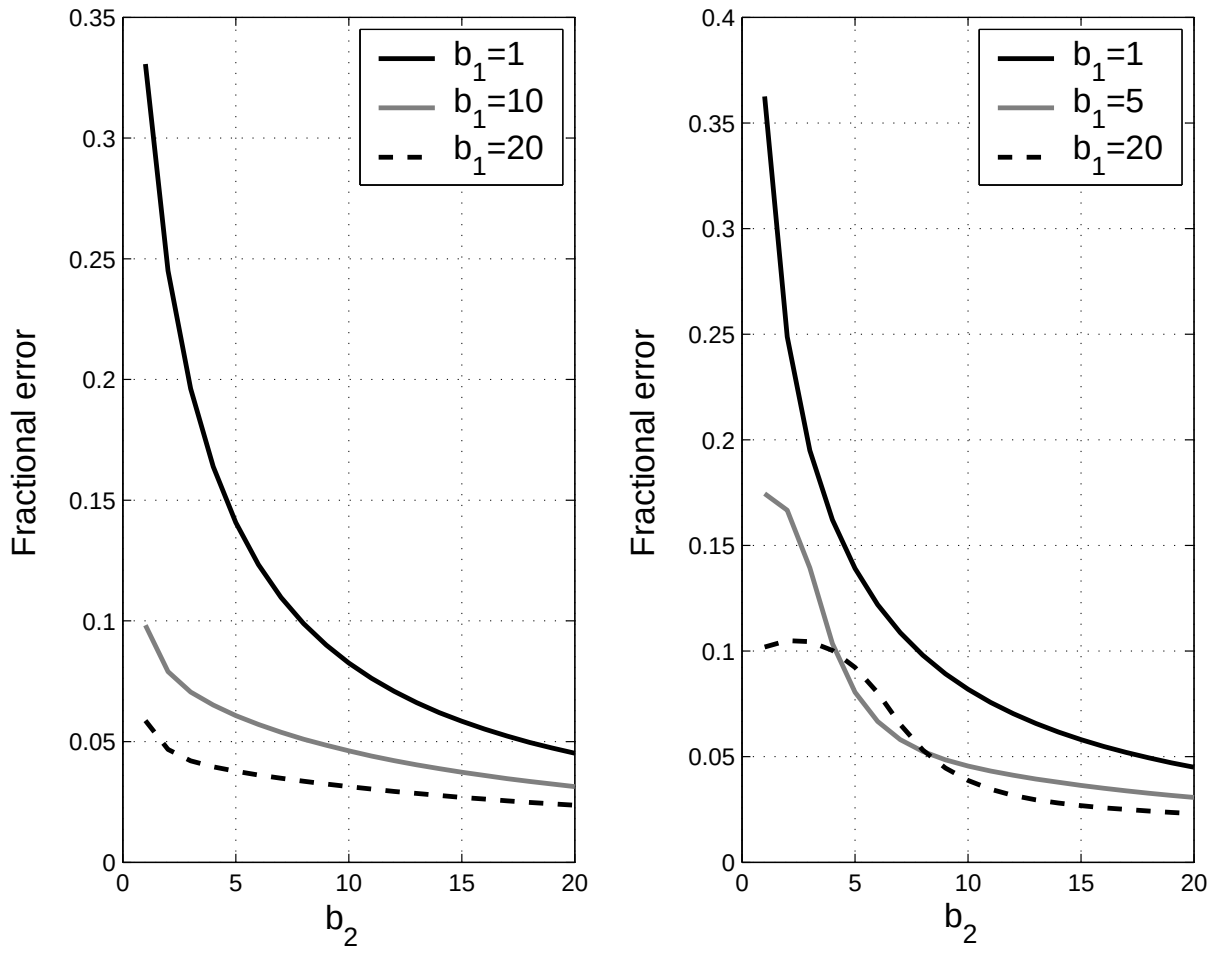

Fig. 13. Numerical example 3